\documentclass[a4paper]{aa}
\usepackage{graphicx}
\usepackage{txfonts} 
\newcommand{\HI}{\ion{H}{i}}
\newcommand{\NII}{\ion{N}{ii}}
\newcommand{\OI}{\ion{O}{i}}
\newcommand{\OII}{\ion{O}{ii}}
\newcommand{\OIII}{\ion{O}{iii}}
\newcommand{\SII}{\ion{S}{ii}}
\newcommand{\Msun}{$M_{\odot}$}

\begin{document}
\title{IC 4200: a gas-rich early-type galaxy formed via a major merger}

   \author{
           P. Serra \inst{1},
           S.C. Trager \inst{1},
           J.M. van der Hulst \inst{1},
           T.A. Oosterloo \inst{1,2}
           R. Morganti \inst{1,2}
           }

   \institute{
              (1) Kapteyn Astronomical Institute, University of Groningen, P.O. Box 800, NL-9700 AV Groningen, the Netherlands \\
              (2) Netherlands Foundation for Research in Astronomy, Postbus 2, NL-7990 AA Dwingeloo, the Netherlands \\
             }

   \date{Received ...; accepted ...}

  \abstract{

 We present the result of radio and optical observations of the S0 galaxy IC 4200. We find that the galaxy hosts 8.5$\times{10^9}$ \Msun\ of \HI\ rotating on a $\sim$90 deg warped disk extended out to 60 kpc from the centre of the galaxy. Optical spectroscopy reveals a simple-stellar-population-equivalent age of 1.5 Gyr in the centre of the galaxy and $V$- and $R$-band images show stellar shells. Ionised gas is observed within the stellar body and is kinematically decoupled from the stars and characterised by LINER-like line ratios.We interpret these observational results as evidence for a major merger origin of IC 4200, and date the merger back to 1-3 Gyr ago.

   }

   \authorrunning{Serra, Trager, Van der Hulst, Oosterloo, Morganti}

   \maketitle

\section{Introduction}

Observations during the last decades have been changing the traditional view of early-type galaxies. Once thought to be old, passively evolving and kinematically-relaxed systems devoid of gas, early-type galaxies have been discovered to host subtle but frequent (hence important) deviations from this simple picture: for example, morphological fine structure, intermediate-age stellar populations and a multi-phase interstellar medium. Explaining how these features come to be is an essential requirement for any theory of galaxy formation and evolution.

Among the observed features, neutral hydrogen gas has been found in a significant fraction of early-type galaxies. The detection rate varies from 5 to 45\% depending on the sample and the depth of the observations (Sadler et al.\ 2002) and seems to be higher in the field (Sadler et al.\ 2001; for individual examples of gas-rich galaxies see Schiminovich et al.\ 1994, 1995, 1997; Oosterloo et al.\ 2002). This \HI\ is usually interpreted in terms of recent accretion of a gas-rich satellite galaxy or equal-mass merger. In particular, major mergers are likely to  be the formation mechanism of the very \HI-rich early-type galaxies observed recently, with extremely massive ($>$10$^9$ \Msun), extended (tens of kpc) and quite regular \HI\ structures surrounding the stellar body (V\'{e}ron-Cetty et al.\ 1995; Morganti et al.\ 1997; Sadler et al.\ 2002; see also the case of NGC 3656 discussed in Balcells et al.\ 2001). Simulations have confirmed that such systems can form during the merger of similarly sized gas-rich galaxies as a result of the re-accretion of high-angular-momentum \HI\ (Hibbard \& Mihos 1995; Barnes 2002). However, other scenarios must be taken into account. Theoretical works showed that galaxies can accrete gas from the IGM via a cold mode (Binney 1977; Fardal et al.\ 2001; Keres et al.\ 2005). This could provide the supply of atomic gas necessary to build \HI\ structures without requiring any galaxy-galaxy interaction. It is not clear which of these processes plays the dominant role in the formation of \HI-rich E/S0's.

Optical observations can shed light on this picture. Unlike accretion from the IGM, merging should produce long-lived signatures in the stellar remnant (Mihos 1995) along with the gas features. During a merger part of the \HI\ of the progenitors is driven toward the centre of the newly formed galaxy, triggering star formation. Moreover, the incomplete relaxation of the stellar body can result in morphological disturbances visible long after the formation event (Hernquist \& Spergel 1992). Such signatures of recent hierarchical assembling have already been observed in the optical images and spectra of early-type galaxies (e.g., Malin \& Carter 1983; Trager et al.\ 2000). There are also suggestions that the morphological disturbances are associated with intermediate-age stellar populations (Schweizer et al.\ 1990; Schweizer \& Seitzer 1992; de Jong \& Davies 1997).

On the basis of these results, an analysis of stellar features in gas-rich E/S0's appears to be a promising way of unveiling their formation history. If these galaxies acquired their \HI\ via mergers rather than accretion from the IGM, stellar and atomic-gas features, being caused by the same event, should be correlated and give the same indications of the particular process that originated each galaxy and of the time passed since then. Despite observations suggesting that galaxies with optical fine structure are more likely to be detected at 21 cm (Van Gorkom \& Schiminovich 1997), other studies indicate that the relation between stellar and \HI\ properties is not straightforward and other factors (for example environment) play a role (Sansom et al.\ 2000, Hibbard \& Sansom 2003). On the other hand, very little is known about the correlation between \HI\ and stellar populations, as galaxies with good \HI\ data have not been well-studied in their stellar content and vice-versa.

To shed light on this picture we have started an observational project aimed to study the \HI-gas content, stellar populations and optical morphology of a sample of gas-rich early-type galaxies. To test the connection between \HI\ gas and stellar populations/morphology, we will also analyse the optical spectra and images of a control sample of gas-poor galaxies.

We note that optical data will also enable us to investigate the properties of the warm gas (10$^4$ K) in early-type galaxies. More than half of the E/S0's show ionised gas emission lines in their optical spectra (Phillips et al.\ 1986; Sarzi et al.\ 2005). The kinematics of this gas is not always consistent with simple coplanar rotation and is often decoupled from the stellar motion (e.g. Falc\'{o}n-Barroso et al.\ 2005), providing clues on the recent history of these systems and on the processes that bring gas into them (Sarzi et al.\ 2005). Linking ionised and atomic gas observations could provide more complete and detailed indications in this sense.

Here we present the results of optical imaging, long-slit spectroscopy and radio observations of the galaxy IC 4200.  This is the first galaxy of the sample for which we obtained both optical and radio data. Our aim is to show the technique that we will apply to all the galaxies of our sample and stress its potential.  The main observational results are a very massive ($8.5\times{10^9}$ \Msun) and extended ($\sim$60 kpc) \HI\ warped disk around the stellar body, an intermediate-age stellar population $\sim$1.5 Gyr old in the centre of the galaxy, optical shells, and ionised gas kinematically decoupled from the stars and characterised by LINER-like line ratios. The combination of these results strongly suggests that IC 4200 formed between 1 and 3 Gyr ago as a result of a major merger rather than by gas accretion from the IGM or accretion of a satellite galaxy.

The paper is structured as follows: in Section 2 the general properties of IC 4200 are discussed; in Section 3 radio, optical spectroscopy and optical imaging observations and data reduction are described; the results of these observations are presented in Section 4; a discussion of the global picture that we derive from these results is given in Section 5; and final conclusions are drawn in Section 6. Appendixes \ref{Lick/IDScal} and \ref{vdispcorr} give details of the calibration of the absorption-line strengths used in the stellar population analysis.  Appendix \ref{eltables} contains two tables that give informations on the kinematics of stars and ionised gas and on the line-strength indices across the galaxy. The tables are available electronically at the Centre de Donn\'{e}es astronomiques de Strasbourg (CDS). 

\section{Basic Properties of IC 4200}
\label{basicprop}

\begin{table}
\begin{center}
\caption{General properties of IC 4200}
\label{generalproperties}
\begin{tabular}{c c c}
\hline
\hline
\noalign{\smallskip}
RA (J2000) &  13h 09m 34.7s & c\\
Dec (J2000) & --51d 58m 07s & c\\
type & S0 & b\\
stellar heliocentric velocity  &  3850 km/s & a\\
\HI\ heliocentric velocity & 3890 km/s & a \\
$z$ & 0.0132 & \\
distance & 55 Mpc & \\
angular scale & 16 kpc/arcmin & \\
$M_B$ & --21.33 & c\\
$L_B$ & 5.30$\times{10^{10}}$L$_{\odot}$ & \\
$\sigma_{\rm stars,centre}$ & 240 km/s & a \\
$M_{\rm \HI}$ & 8.54$\times{10^9}$\Msun & a\\
$F_{\rm 60 \mu m}$ & 0.17 Jy & d\\
F$_{\rm 1.4 GHz}$ & 11.5 mJy & a \\
\noalign{\smallskip}
\hline
\end{tabular}
\end{center}
References: (a) this paper, (b) de Vaucouleurs et al.\ (1991), (c) HyperLeda, (d) Thronson et al.\ 1989.
\end{table}

Table \ref{generalproperties} summarises the main properties of IC 4200. The galaxy is a high-luminosity S0. Its heliocentric stellar systemic velocity derived from our optical long-slit spectroscopy is 3850 km/s, which corresponds to a distance of 55 Mpc and a scale of 16 kpc/arcmin \footnote{$H_0$ is assumed to be 70 km s$^{-1}$ Mpc$^{-1}$ throughout this paper.}. The heliocentric systemic velocity of the \HI\ gas is $\sim3890$ km/s \footnote{This is the systemic velocity of the gas model that reproduces our 21-cm wavelength observations; the model is described in Sect.\ref{higasres}.}.

IC 4200 was first detected in \HI\ by the HIPASS\footnote{\HI\ Parkes All Sky Survey.} survey (Meyer et al.\ 2004). \HI\ emission was measured in the velocity range 3647-4129 km/s resulting in a total flux of 13.3 Jy$\times$km/s within a beam of 14.3 arcmin FWHM (230 kpc). This detection was confirmed by Oosterloo et al.\ (2006, in preparation) with the ATCA (Australian Telescope Compact Array), as a part of follow-up observations of $\sim$40 early-type galaxies detected by HIPASS. To establish how many of these detections are caused by confusion with independent gas clouds, they observed each galaxy for three hours. They found that in roughly half of the cases \HI\ gas is associated to the early-type galaxy. IC 4200 is one of these: a very extended gas disk was observed around the stellar body and its \HI\ mass was estimated as 5.9$\times{10^9}$ \Msun. On the basis of these observations, we included IC 4200 in our sample for the study of the relation between stellar and \HI\ properties in early-type galaxies.

We characterised the environment of IC 4200 by looking for extragalactic objects within 100 arcmin ($\sim$1.6 Mpc at this redshift) and 300 km/s from the galaxy; for this search we used the NED database. The only object with measured redshift that we found is ESO 219-20, an Sc galaxy at a projected distance of 83 arcmin (1.3 Mpc) from IC 4200 and at a velocity of 3958 km/s. Within 20 arcmin (320 kpc), we found eleven sources without measured redshift, ten of which appear only in the catalogue of 2MASS\footnote{2 Micron All Sky Survey.} detections. All of them have small optical size compared to IC 4200. Using the \HI\ image derived from our radio observations (Sect.\ref{hisection}), we found that only one of the 2MASS objects, 2MASX J13090029--5153544, contains detectable atomic gas at the same velocity of IC 4200: this is an Sc galaxy 6.8 arcmin (110 kpc) northwest of IC 4200 and is the closest object to our target. In addition, our \HI\ image shows two more sources at $\sim$3900 km/s. These are 15 and 17 arcmin (240 and 270 kpc) distant from IC 4200, but we could not identify their optical counterparts with the available catalogues and DSS\footnote{Digitised Sky Survey.} images. Although we cannot conclude that IC 4200 is an isolated galaxy, it appears that the few possible nearby galaxies are much less massive than IC 4200 itself.

\begin{table}
\begin{center}
\caption{ATCA configurations}
\label{baselines}
\begin{tabular}{c c c c}
\hline
\hline
\noalign{\smallskip}
 & & Baselines & \\ \vspace{0.2cm}
Dates & Configuration & range(m) & Integration \\
5,6 March 2004 & 750A & 77-735 & 2$\times{12}$ h \\
11 April 2004 & EW367 & 46-367 & 12 h \\
\noalign{\smallskip}
\hline
\end{tabular}
\end{center}
\end{table}

\begin{table}
\begin{center}
\caption{ATCA observations}
\label{atca}
\begin{tabular}{c c}
\hline
\hline
\noalign{\smallskip}

Centre RA (J2000) & 13:09:34.699 \\
Centre Dec (J2000) & --51:58:06.99 \\
Frequency & 1402 MHz\\
Bandwidth & 16 MHz \\
Number of channels & 512 \\
Velocity resolution$^1$ & 13 km/s \\
 
\noalign{\smallskip} \hline \noalign{\smallskip}

\emph{21-cm emission} & \emph{(natural weighting)}\\
Synthesised beam & 78"$\times{67"}$\\
 & $PA$=6.8 deg\\
RMS noise  & 0.9 mJy/beam \\
Maximum signal & 10.5 mJy/beam \\

\noalign{\smallskip} \hline \noalign{\smallskip}

\emph{1.4 GHz continuum} & \emph{(robust=0.5)}\\
Synthesised beam & 68"$\times{59"}$\\
 & $PA$=5.3 deg\\
RMS noise & 0.24 mJy/beam\\
Maximum signal & 37.21 mJy/beam \\

\noalign{\smallskip}
\hline
\end{tabular}
\end{center}
(1) After Hanning smoothing.
\end{table}

IC 4200 has been observed by many authors at different wavelengths. Malin \& Carter (1983) included it in their catalogue of shell galaxies identified by un-sharp masking. According to these authors, shells are observable in the southern part of the galaxy optical image. Carter et al.\ (1988) followed up this study by observing the optical spectra of a large fraction of the shell galaxies. They measured an intense [\OII] emission line in the spectrum of IC 4200.

$UBVRI$ aperture photometry was obtained by Winkler (1997) as a part of his study of a sample of Seyfert galaxies. Within a 20-arcsec-diameter aperture IC 4200 was found to have $V$=13.82, $U-B$=0.77, $B-V$=1.33 and $V-R$=0.78 before correction for the galactic extinction $E(B-V)$=0.30.

Thronson et al.\ (1989) studied the mid- and far-infrared emission from early-type shell galaxies; they measured a flux density of $(170\pm30)$ mJy at 60 $\mu$m from IC 4200. A flux density of 0.24 Jy was measured at 100 $\mu$m but this measure was flagged as questionable. Finally, IC 4200 has not been detected by any X-ray survey.

\section{Observations and Data Reduction}
\subsection{Radio}

To study the morphology and kinematics of the \HI\ gas around IC 4200, radio observations were carried out with the ATCA in March and April 2004. Our aim is to obtain resolution high enough to study the velocity structure of the gas, as well as to detect faint extended emission. To achieve this, the galaxy was observed for 3$\times{12}$ h combining the configurations 750A (2$\times{12}$ h) and the very compact EW367 (12 h). Details of the observations can be found in Table \ref{baselines} and \ref{atca}.

In all three 12-h observing runs, we used a 16-MHz bandwidth with 512 channels centred on a frequency of 1402 MHz. PKS 0823--500 and PKS 1934--638 were observed respectively at the beginning and end of each run to determine the bandpass and the flux-density scale; gain changes were monitored by observing the secondary calibrator PKS 1215--457 for 3 minutes for every 40 minutes integration on IC 4200.

Data were reduced using MIRAD software (Sault et al.\ 1995) in a standard way.  1.4 GHz continuum and 21-cm emission line were separated by continuum linear fitting through the line-free channels in the UV plane.

\subsubsection{\HI}
\label{hisection}

The final \HI\ emission data cube was obtained by combining the continuum-free data from each run and inverting with natural weighting to maximise the sensitivity to low-column-density, extended regions. Hanning smoothing was applied resulting in a velocity resolution of 13 km/s. The cube consists of 110 7-km/s-wide channels and is characterised by a rms noise of $\sim$0.9 mJy/beam in each channel. The beam has axes of 78 and 67 arcsec ($PA$=6.8 deg) which  at the redshift of IC 4200 correspond to 20.8 and 17.8 kpc respectively.

\HI\ gas is detected around IC 4200 in the velocity range 3590-4130 km/s, consistent with HIPASS detection except for very faint emission at the lower velocities. 21-cm line emission is detected also NW of IC 4200 between 3670 and 3900 km/s; the latter is associated to the 6.8 arcmin (110 kpc) distant galaxy 2MASX J13090029-5153544, which we will call IC 4200-A\footnote{As mentioned in Sect.\ref{basicprop}, 2MASX J13090029--5153544 appears only in the catalogue of the 2MASS survey and is the closest object to IC 4200.}.

The total \HI\ image was built using a masked version of the data cube; the mask was built to select pixels above 3 $\sigma$ in either the 60-arcsec or the 180-arcsec smoothed versions of the data cube and above 2 $\sigma$ in the original cube. The \HI\ masses detected around IC 4200 and IC 4200-A are respectively $8.54\times{10^9}$\Msun\ and $7.3\times{10^8}$ \Msun\ and extend out to 60 kpc and 22 kpc from their centres. As mentioned in Sect.\ref{basicprop}, less massive \HI\ systems without any visible optical counterpart are detected in the same velocity range $\sim$17 arcmin (270 kpc) south and $\sim$15 arcmin (240 kpc) west of IC 4200. The \HI\ masses of these systems are 1.5 and $1.7\times{10^8}$ \Msun.

\subsubsection{Continuum}

The 1.4-GHz continuum image was made by combining the 21-cm emission-free data from each observing run and inverting them with $robust$=0.5 (for details on the $robust$ parameter see Briggs 1995). The beam of the image has axes of 68 and 59 arcsec ($PA$=5.3 deg). The rms noise is of 0.24 mJy/beam. A flux density of 11.5 mJy is detected within one beam centred on IC 4200. IC 4200-A is also detected at a 6 $\sigma$ level, with a total unresolved flux of 1.5 mJy.

\subsection{Optical Spectroscopy}
\label{genoptspec}

\begin{table}
\begin{center}
\caption{EMMI long-slit spectroscopy}
\label{spec}
\begin{tabular}{c c}
\hline
\hline
\noalign{\smallskip}
Date & 12 July 2004 \\
Period & ESO-073.A \\
Telescope & NTT(3.6 m) \\
Instrument & EMMI (red arm) \\
Dispersing medium & Grism 5 \\
Spectral range & 3800-7020 \AA \\
Dispersion & 1.66 \AA/pixel \\
Resolution & 5 \AA \ FWHM \\
Detector & $2\times{2048}\times{4096}$ MIT/LL CCD \\\
Binning & 2$\times{2}$ \\
Gain & 1.32 e$^-$/ADU \\
Read-out-noise & 3.825 e$^-$\\
Scale & 0.332 arcsec/pixel \\
Slit length & 8 arcmin\\
Slit width & 1 and 5 arcsec \\
\noalign{\smallskip}
\hline
\end{tabular}
\end{center}
Dispersion, scale, gain and $RON$ refer to 2$\times{2}$ binning.
\end{table}

Long-slit spectroscopy of IC 4200 was carried out on 12 July 2004 at ESO La Silla Observatory using EMMI mounted on the NTT. The main purpose of these observations is to study the stellar content and kinematics of the galaxy. To achieve this target our spectra cover the range 4000 to 7000 \AA \ with dispersion 1.66 \AA /pixel (with 2$\times{2}$ binning) and resolution of $\sim$5 \AA. Observational details can be found in Table \ref{spec}.

The time spent on the galaxy was divided between two perpendicular slit positions along the optical major and minor axes (respectively at $PA$=150 and 60 deg). Because of bad weather, the total exposure times are 3$\times{1350}$ sec along the major axis and only 2$\times{1350}$ sec along minor axis. The galaxy was observed through a 1-arcsec-wide and 8-arcmin-long slit with an average seeing of $\sim$1.7 arcsec. During the same observing run,  we observed the spectra of 3 spectrophotometric standard stars for flux calibration purposes through a 5-arcsec-wide slit: the observed stars are EG 21, LTT 4816 and LTT 6248. Eight more stars (which we will call Lick/IDS stars) were observed through the 1-arcsec-wide slit to calibrate our line-strength indices to the Lick/IDS system (Worthey et al.\ 1994) in order to study the stellar populations; Table \ref{LICKlist} lists the stars with their spectral type and exposure times.

\begin{table}
\begin{center}
\caption{Lick/IDS stars}
\label{LICKlist}
\begin{tabular}{c c c c}
\hline
\hline
\noalign{\smallskip}
Star & Spectral & Stellar & Exposure\\
& Type & Library & Time (sec) \\
\noalign{\smallskip}
\hline
\noalign{\smallskip}
HD 114113 & K3III & INDO-US & 42\\
HD 124850 & F7IV & MILES, INDO-US & 40 \\
HD 125454 & G8III & INDO-US,Lick/IDS & 120\\
HD 126218 & K0III & MILES & 120 \\
HD 131977 & K4V & MILES,Lick/IDS & 240\\
HD 137052 & F5IV & INDO-US & 120\\
HD 138716 & K1IV & INDO-US & 120 \\
HD 139446 & G8III/IV & MILES,INDO-US & 130\\
\noalign{\smallskip}
\hline
\end{tabular}
\end{center}
\end{table}

\begin{table}
\begin{center}
\caption{Lick/IDS index definitions $^1$}
\label{ind_def}
\begin{tabular}{c c c c}
\hline
\hline
\noalign{\smallskip}
Index & Index band & Blue continuum & Red continuum \\
\noalign{\smallskip}
\hline
\noalign{\smallskip}
H$\beta$    & 4847.875 & 4827.875 & 4876.625 \\
            & 4876.625 & 4847.875 & 4891.625 \\
Mg$b$       & 5160.125 & 5142.625 & 5191.375 \\
            & 5192.625 & 5161.375 & 5206.375 \\
Fe5270      & 5245.650 & 5233.150 & 5285.650 \\
            & 5285.650 & 5248.150 & 5318.150 \\
Fe5335      & 5312.125 & 5304.625 & 5353.375 \\
            & 5352.125 & 5315.875 & 5363.375 \\
H$\gamma_{\rm A}$ & 4319.750 & 4283.500 & 4367.250 \\
            & 4363.500 & 4319.750 & 4419.750 \\
H$\delta_{\rm A}$ & 4083.500 & 4041.600 & 4128.500 \\
            & 4122.250 & 4079.750 & 4161.000 \\
H$\gamma_{\rm F}$ & 4331.250 & 4283.500 & 4354.750 \\
            & 4352.250 & 4319.750 & 4384.750 \\
H$\delta_{\rm F}$ & 4091.000 & 4057.250 & 4114.750 \\
            & 4112.250 & 4088.500 & 4137.250 \\
\noalign{\smallskip}
\hline
\end{tabular}
\end{center}
(1) Wavelengths are in angstroms.
\end{table}

\begin{table}
\begin{center}
\caption{Calibration onto the Lick/IDS system}
\label{Lick_cal}
\begin{tabular}{c c c c}
\hline
\hline
\noalign{\smallskip}
Index    & Slope ($m$) & Intercept ($q$) & $\chi^2$ \\
\noalign{\smallskip}
\hline
\noalign{\smallskip}
H$\beta$    & 1.05 & --0.23 &  1.7 \\
Mg$b$       & 1.15 & --0.16 & 12.6 \\
Fe5270      & 1.03 & --0.12 & 22.7 \\
Fe5335      & 0.77 &   0.40 &  0.8 \\
H$\gamma_{\rm A}$ & 1.08 & --1.45 &  7.5 \\
H$\delta_{\rm A}$ & 1.03 & --1.09 &  1.4 \\
H$\gamma_{\rm F}$ & 1.00 & --0.51 &  3.3 \\
H$\delta_{\rm F}$ & 1.05 & --0.45 &  2.5 \\
\noalign{\smallskip}
\hline
\end{tabular}
\end{center}
The measured value of a given index $i$ is: $i_{\rm meas}=m\cdot i_{\rm ref}+q$, where $i_{\rm ref}$ is the value on the Lick/IDS system.
\end{table}

Data were reduced using a variety of suitable programs written by Daniel Kelson, Scott Trager and Paolo Serra in the Python programming language. Reduction was performed in a standard way. The rms of the wavelength calibration of the frames is on average 0.05 \AA \ and never exceeds 0.1 \AA. Sky emission was modelled and subtracted before re-binning the frames to correct for distortion and wavelength calibration; this allows one to better sample the sky lines profiles by exploiting the distortion and therefore reduces the residuals at the line edges (Kelson 2003).

One-dimensional spectra of the calibration stars were extracted. A flux calibration curve was derived from the spectrophotometric standard stars and applied to the Lick/IDS stars. The latter were then used to calibrate the measured line-strength indices to the Lick/IDS system. Each index was calibrated by determining the linear relation between its measured value and the one on the Lick/IDS system (see Appendix \ref{Lick/IDScal} for details). To efficiently disentangle age from metallicity in the spectrum of the galaxy, we use the indices H$\beta$, Mg$b$, Fe5270 and Fe5335. We test the result against indices of higher-order Balmer-lines: H$\gamma_{\rm A}$, H$\delta_{\rm A}$, H$\gamma_{\rm F}$, and H$\delta_{\rm F}$, defined in Worthey \& Ottaviani (1997). All the indices used are defined in Table \ref{ind_def}. Table \ref{Lick_cal} contains the coefficients of the relations that calibrate these indices to the Lick/IDS system.

The flux calibration was finally applied to the two-dimensional spectra along the major and minor axes of IC 4200. Any information of interest can be derived from these calibrated 2D spectra as a function of the radial coordinate. To do so we built 1D spectral bins of appropriate S/N. Higher S/N is needed for line-strength indices measurement than for kinematics analysis in order to obtain reasonable errors; therefore, we extracted bins of S/N$\sim$30/pixel and of S/N$\sim$80/pixel at 5000 \AA \ for these two different purposes. In the rest of the paper low- and high-S/N 1D bins will be referred to as $B_{30}$'s and $B_{80}$'s.

\subsubsection{Disentangling stars from ionised gas}
\label{fit}

\begin{figure}
\includegraphics[width=8.5cm,angle=0]{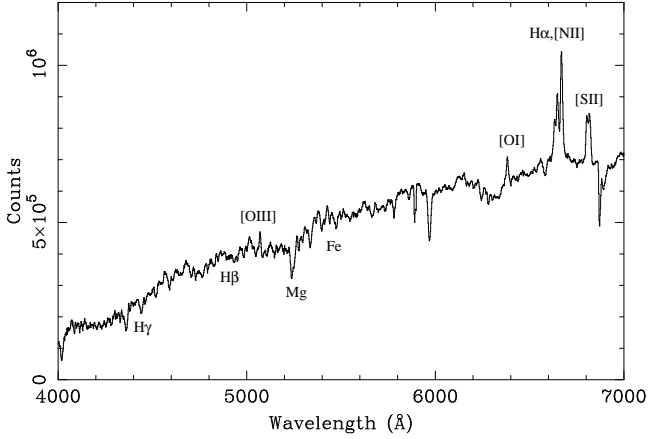}
\caption{Central spectrum $B_{80}$ of IC 4200 along the optical major axis in arbitrary flux units. Ionised gas emission lines and the main stellar absorption features are highlighted.}
\label{B80MJcen}
\end{figure}

Fig.\ref{B80MJcen} shows the central $B_{80}$ along the major axis. [\OI], [\OIII], [\NII], H$\alpha$ and [\SII] emission lines are easily recognised and the H$\beta$ absorption feature is completely filled by emission; even $H\gamma$ absorption is contaminated by the ionised gas. The same features are observed in all the $B_{80}$'s. On one hand emission lines enable us to study the physical processes and kinematics of the gas within the stellar body. However, they make the stellar population analysis more complicated by contaminating or even hiding important absorption features (e.g. H$\beta$, crucial for determining the age of stellar populations). Before measuring line-strength indices from the $B_{80}$'s it is therefore necessary to remove the ionised gas emission from these spectra. We separate gas and stellar spectra also along the $B_{30}$'s to study the kinematics of the gas and of the stars separately.

To disentangle the ionised gas and the stellar contribution to the spectrum of IC 4200, we fit Bruzual \& Charlot (2003, hereafter BC03) stellar population synthesis models to the $B_{80}$'s and $B_{30}$'s in their emission-free wavelength ranges. We use the high-resolution (3 \AA) models built with the Padova 1994-isochrones and a Salpeter IMF, selecting metallicities $Z$=0.004, 0.008, 0.02, 0.05 ($Z_{\odot}$=0.02) and ages between 1 and 20 Gyr, evenly spaced in logarithmic steps of 0.1. We apply the direct fitting method described in Kelson et al.\ (2000). In this method the model spectra, smoothed to the instrumental resolution of our data, are convolved with a parameterised line-of-sight velocity distribution (LOSVD); this is assumed to be a Gaussian and has two free parameters, the velocity $v$ and the velocity dispersion $\sigma$. Continuum matching is obtained by an additive and a multiplicative correction of the model spectra, both dependent on wavelength via polynomial functions. The former is necessary because the overall shape of the galaxy spectrum might be not well represented by the models; the latter removes any mismatch caused by the instrumental response and provides the necessary normalisation.  We have tested the effect of using the additive correction during the fitting procedure. We found that while applying this correction gives a better quality of the fit, it does not change the best fitting spectrum within 1.5 $\sigma$ at any wavelength. 

A best fitting model $M_k$ is associated to each bin $B_k$. To separate gas emission from the underlying stellar absorption spectrum we use the technique described in Emsellem et al.\ (2002). The residual $G_k=B_k-M_k$ represents an estimate to the emission from ionised gas. We then build a model emission spectrum $\widetilde{G_k}$ by fitting each emission line in $G_k$ with a Gaussian curve. The model emission spectrum is finally subtracted from the original bin, leaving us with a pure stellar absorption spectrum $S_k=B_k-\widetilde{G_k}$.

As a part of the fitting procedure of the $B_{30}$'s we obtain the stellar LOSVD along the two axis. Furthermore, ionised gas velocities are determined from the central wavelengths of the single-Gaussian fit to the emission lines performed to build the $\widetilde{G_{30}}$'s. Finally, we use the flux of the ionised gas emission lines -- measured when modelling the $G_{80}$'s -- to examine the ionisation mechanism in the galaxy.  Informations on the stellar and ionised-gas kinematics along the $B_{30}$'s are listed in Table C.1.

\subsubsection{Absorption-line indices}
\label{indices}

We use the $S_{80}$'s to study the stellar populations of IC 4200. For each bin $S_k$ five main steps have to be performed in order to properly determine the Simple-Stellar-Populations-equivalent (SSP-equivalent) age, metallicity and abundance ratio:

\begin{itemize}
\item broaden $S_k$ to match the Lick/IDS resolution
\item measure the indices $i_n$ ($n=1,...,N_{\rm indices}$)
\item correct the indices for the effect of the stellar velocity dispersion $\sigma_k$
\item bring the corrected indices to the Lick/IDS system
\item compare the final values of the indices to the models
\end{itemize}

The first two steps are performed by feeding the stellar velocity $v_k$ to a Python code that broadens $S_k$ to the Lick/IDS resolution (Worthey \& Ottaviani 1997) and measures the indices placing the bandpasses of Table \ref{ind_def} in the appropriate wavelength range according to the stellar velocity.

Velocity dispersion broadens the spectral feature of galaxies with respect to the ones of stellar spectra, altering the line strengths by transporting flux in and out of the bandpasses that define each index. Therefore, before comparing the measured indices to the models (which are built with zero velocity dispersion) we need to compensate for this effect. To perform this correction we use the result of BC03-models fitting of Sect.\ref{fit}. We measure each index $i$ from the best fitting model $M_k$ and from its zero-velocity-dispersion equivalent $M_{0,k}$, obtaining $i_{k}$ and $i_{0,k}$ respectively. We then calculate $f=i_{k}/i_{0,k}$. Finally, we divide the index measured from $S_k$ by $f$, so correcting it for the effect of the stellar velocity dispersion in $S_k$. Plots of the correction coefficient $f$ as a function of the velocity dispersion are given for each index used in this paper in Appendix \ref{vdispcorr}.

The linear transformations that bring the measured indices onto the Lick/IDS system (i.e. the inverse of the transformations of Table \ref{Lick_cal}) are then applied. Table C.2 lists the value of all the Lick/IDS indices fully corrected and calibrated. The result of the comparison of the index values to the models is showed in Sect.\ref{spopandfrac}.

\subsection{Optical Imaging}
\label{optobs}

$V$ and $R$ broad-band imaging of IC 4200 was obtained on 4 February 2005 at La Silla Observatory, ESO. As for the optical spectroscopy, we used the red arm of EMMI on the NTT which is characterised by a field of view of 22.7$\times$22.7 arcmin$^2$. Detector parameters are the same of table \ref{spec} (2$\times{2}$ binning was applied). During the observations the seeing was on average $\sim$0.8 arcsec.

The galaxy was observed for 12$\times{50}$ sec and 12$\times{25}$ sec respectively in $V$ and $R$ band. Each frame was bias subtracted. Flat fielding was performed using $V$ and $R$ median sky flats obtained from multiple twilight exposures. The sky emission was assumed to be equal to the mode of the pixel distribution in each science frame and then subtracted. The calibration onto the magnitude scale was obtained by use of five and seven standard stars in $V$ and $R$ band respectively.

To analyse the image of IC 4200 without contamination from the stars in the field, we built a stellar mask in each band. This was done by applying a 41$\times{41}$ pixel$^2$ median filter that smooths away the stars, dividing the original image by the smoothed one and imposing a threshold on the resulting frame to find the pixels that belong to the stars. We used the mask when measuring the total flux of IC 4200 in the two observed bands. Total $V$ and $R$ magnitude and the $V-R$ colour are $V$=12.65, $R$=11.86 and $V-R$=0.79. The colour is in good agreement with the 0.77 found by Winkler (1997); on the contrary, our magnitude is brighter, but this is a consequence of his narrow aperture \footnote{Indeed, the $V$-magnitude measured within a 20-arcsec-diameter aperture is in agreement with the result of Winkler (1997).}. We corrected magnitudes and $V-R$ colour for galactic reddening using the corrections 1.06 mag in $V$ and 0.85 mag in $R$ given by NED. Corrected magnitudes and colour are $V$=11.59, $R$=11.01, $V-R$=0.58.

\section{Results}

\subsection{\HI}
\label{higasres}

\begin{figure}
\includegraphics[width=8cm]{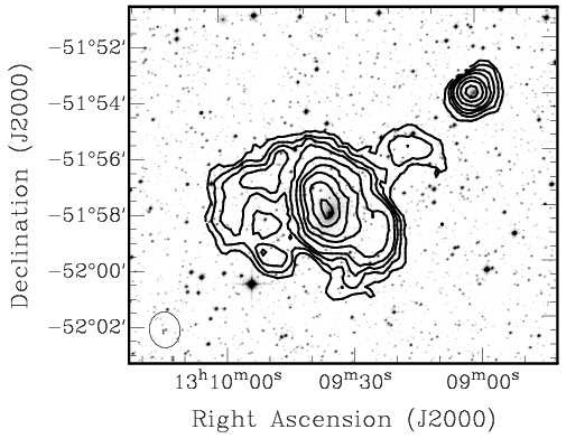}
\caption{Total \HI\ contours on top of the DSS optical image of IC 4200. Contour levels are (0.2, 0.4, 0.6, 0.8, 1.2, 1.6, 2.0, 2.6, 3.2)$\times{10^{20}}$ cm$^{-2}$. The maximum column density is 3.32$\times{10^{20}}$ cm$^{-2}$. The beam has axes of 78 and 67 arcsec ($PA$=6.8 deg). For a grey scale representation of the total \HI\ image see Fig.\ref{model}.}
\label{hitot}
\end{figure}

\begin{figure}
\includegraphics[width=9cm]{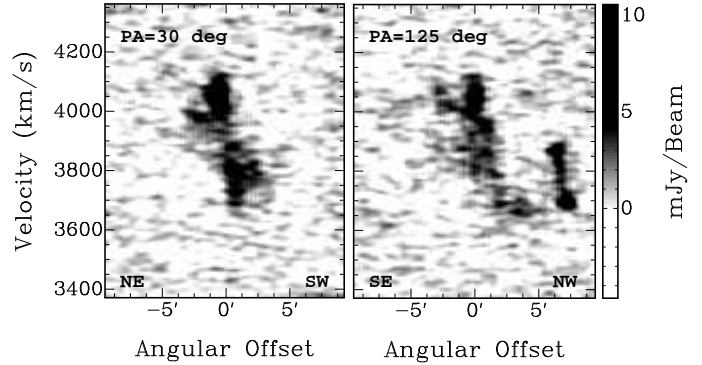}
\caption{Position-velocity diagrams along $PA$=30 deg (left) and $PA$=125 deg (right). The left diagram is aligned along the kinematic major axis of the inner gas region. The right diagram is only approximately aligned along the kinematic major axis of the outer gas region ($PA$=120 deg) to better show the \HI\ connection between IC 4200 and IC 4200-A (lower right; see the discussion in the text).}
\label{pa_mos}
\end{figure}

Fig.\ref{hitot} shows the total \HI\ emission image derived from our ATCA observations. The resolution of the data is low: the whole gas structure around IC 4200 is sampled by only 6 beams along its maximum extension; the one around IC 4200-A is contained within 2 beams. Nevertheless, it is possible to extract a significant amount of useful information.

We detect $8.54\times{10^9}$ \Msun\ of \HI\ gas around IC 4200. The gas goes as far as 60 kpc from the centre of the galaxy and shows a fairly regular configuration. The major axis of the isophotes rotates from $\sim$20 to $\sim$80 deg when  moving out from the centre. The isophotes are quite regular and symmetric around the centre in the inner $\sim$3 beams. The regular inner appearance vanishes in the asymmetric outer regions, where the east side is more extended than the west one and presents a feature that could be a spiral arm. On the northwest the structure is elongated toward the companion.

The velocity structure of the cube shows a certain order, indicating that the gas must have already been in place for some time.  Position-velocity diagrams (Fig.\ref{pa_mos}) suggest that the kinematic major axis is aligned along $\sim$30 deg in the inner region and along $\sim$120 deg in the outer part of the gas structure, following the trend of the isophotal major axis with radius. The kinematic major axis of the outer regions is therefore parallel to the photometric minor axis of the inner region, indicating that this is not a regular disk. The strong rotation of the kinematic major axis with radius also rules out simple coplanar rotation as a model for the motion of the \HI\ gas around IC 4200. 

\begin{figure}
\includegraphics[width=9cm]{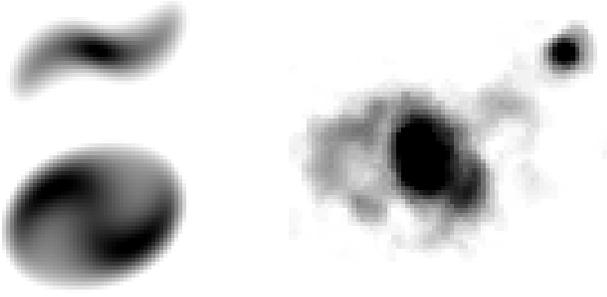}
\caption{On the left: edge-on (top) and sky-projected (bottom) view of the warped gas-disk model. The model is not meant to reproduce the gas density distribution of the \HI\ in IC 4200 but rather its velocity structure (see Fig.\ref{modelandcube}). However, the figure shows that projection effects caused by the warp can produce a spiral pattern like the one observed in the total-\HI\ image of the galaxy, showed on the right.}
\label{model}
\end{figure}

We model the observed \HI\ system in a simple way, finding that a strongly-warped disk satisfactorily reproduces the velocity structure of the cube. The model has a radius of 4 arcmin ($\sim$60  kpc) and a flat rotation curve with $v_{rot}=320$ km/s. We keep the density constant with radius as we are interested in reproducing the kinematics of the disk rather than its detailed density distribution. The inclination is also kept constant as a function of the radius and equal to 45 deg; finally, the major axis $PA$ increases linearly from 30 deg in the centre to 120 deg at the outer radius of 4 arcmin, producing a 90 deg warp.

\begin{figure}
\includegraphics[width=8.75cm]{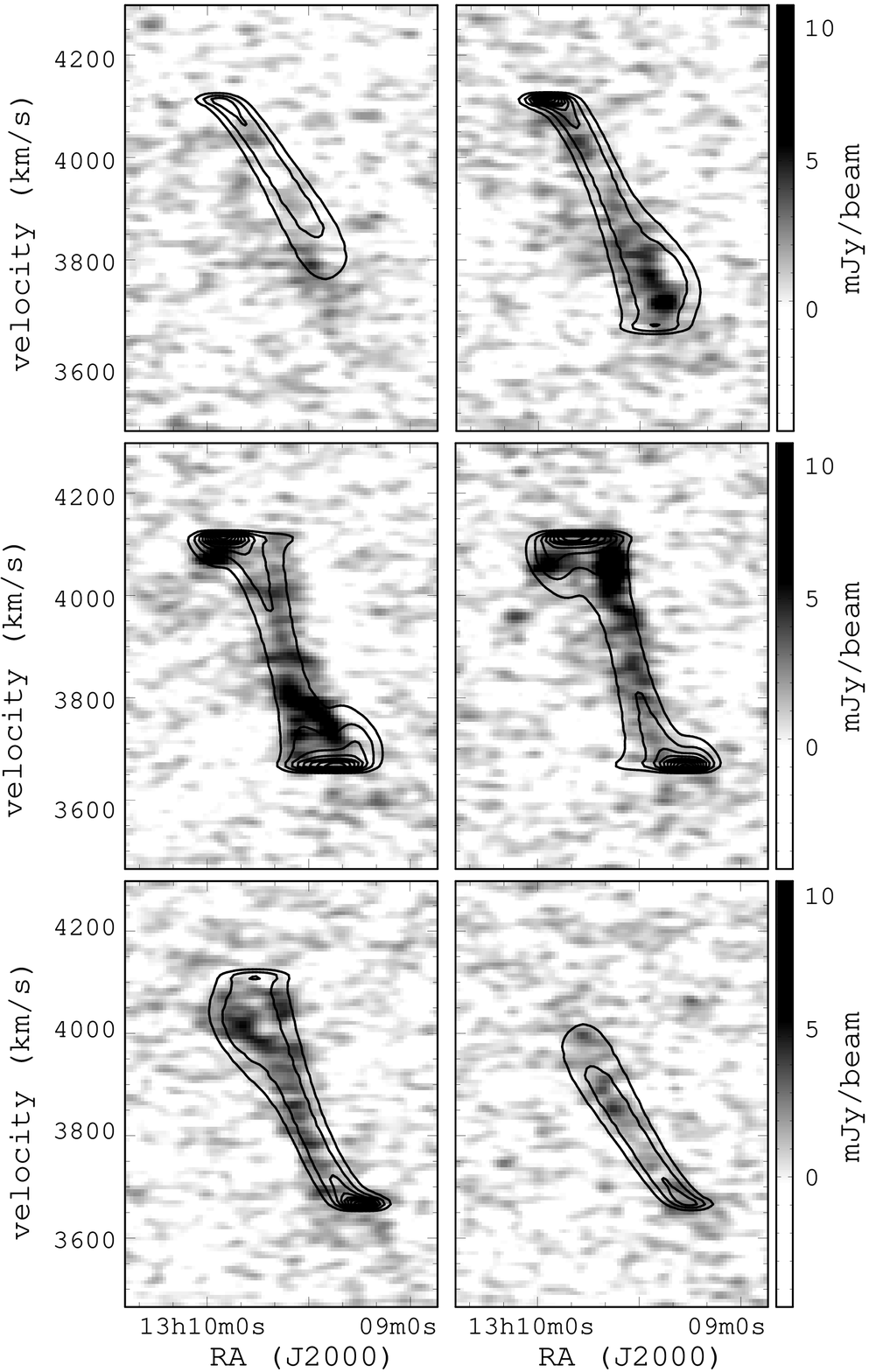}
\includegraphics[width=8.75cm]{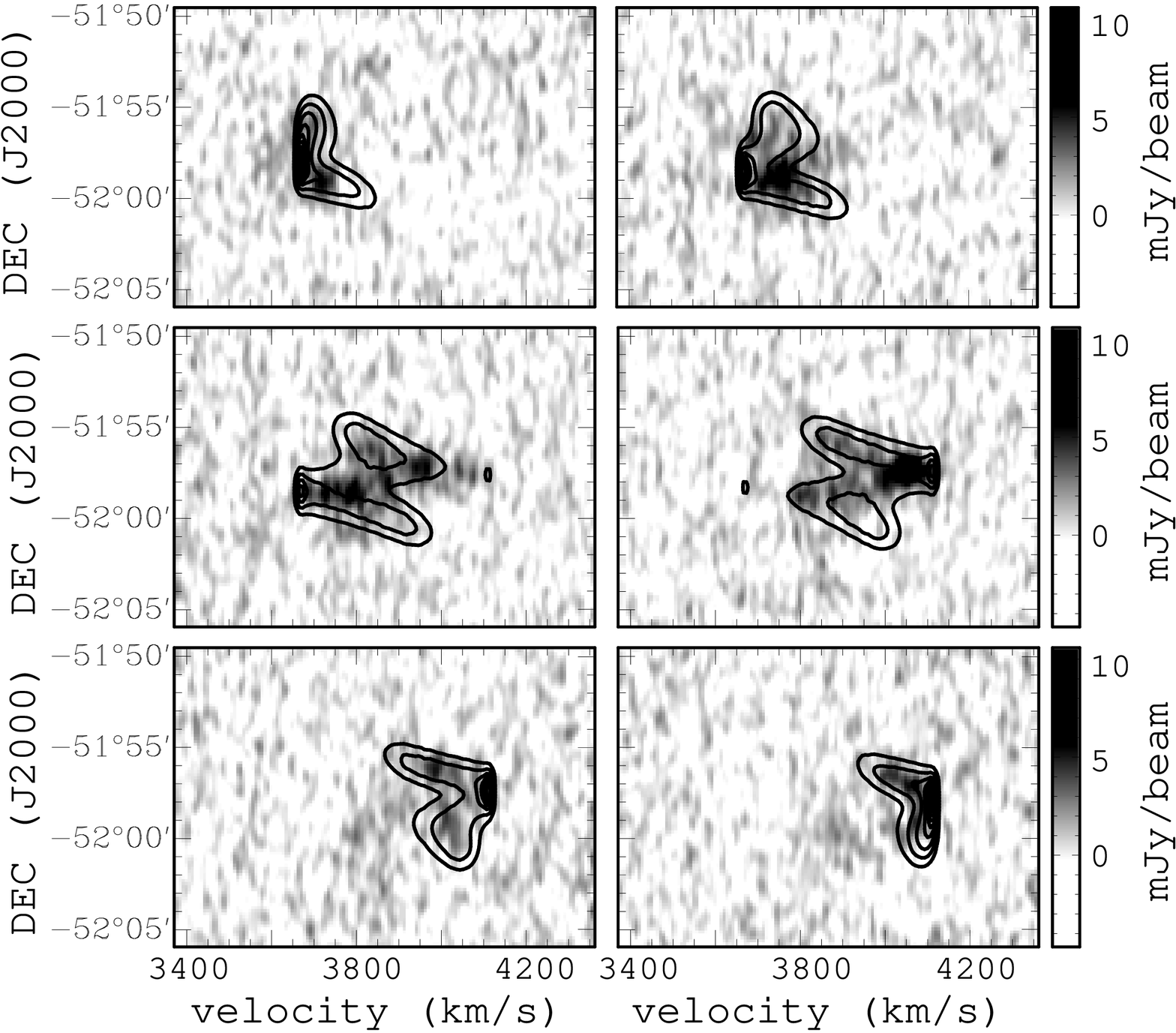}
\caption{Constant-Dec (top six panels) and -RA (bottom six panels) position-velocity diagrams of the observed (grey scale) and modelled (contours) gas system. Both Dec and RA increase from left to right, top to bottom with increments of 1 arcmin.}
\label{modelandcube}
\end{figure}

In Fig.\ref{model} the disk model is shown edge-on and projected on the sky. The strong warp produces projection effects that can simulate spiral arms like the observed one. Fig.\ref{modelandcube} shows the good agreement between data cube and model when looking at constant-RA and -Dec slices. Although the model is not the result of a best fit, only minor changes are allowed in order to reproduce the observed velocity structure. Finely tuning the model parameters is not worth the effort given the resolution of our data. We can therefore conclude that the gas orbits around IC 4200 forming a $\sim$90 deg warped disk with a flat rotation curve with $v_{rot}$=320 km/s. This is the main conclusion of our 21-cm line emission observations.

The \HI\ content of IC 4200 is very high, 1.7 times the one of the Milky Way \footnote{$M_{\rm \HI,MW}=5\times{10^{9}}$ \Msun\ (Henderson et al 1982).}. However, the gas is spread over an extremely wide area; it is therefore very dilute, its column density being below a few times 10$^{20}$ cm$^{-2}$ across most of the system. As a result, no star formation is expected on the scale of the beam size. To quantify this point we estimate the critical surface density for star formation proposed by Kennicutt (1989). Star formation occurs at densities higher than
\begin{displaymath}
\Sigma_{\rm critical}(R)=6.2 \frac{\sigma(\rm km/s) \ v(km/s)}{R(\rm kpc)} 10^{18}\rm cm^{-2},
\end{displaymath}
where $v$ and $\sigma$ are the rotational velocity and the local velocity dispersion of the gas at the radius $R$, and flat rotation has been assumed. Using the model rotational velocity of 320 km/s and a standard velocity dispersion of 7 km/s, the critical column density ranges from 2.5$\times{10^{20}}$ cm$^{-2}$ at the outer edge of the disk to $\sim$10$^{21}$ cm$^{-2}$ in the centre. These values are always higher than the measured column density, which ranges from 2.0$\times{10^{19}}$ to 3.3$\times{10^{20}}$ cm$^{-2}$. Therefore, no star formation is expected on a scale of the order of the beam size. The same conclusion is reached using the criterion proposed by Schaye (2004), where the cooling of the gas triggers gravitational instability and therefore star formation at a fixed surface density of $\sim$6$\times{10^{20}}$ cm$^{-2}$. However, because of the large beam smearing, we cannot exclude that star formation is occurring in unresolved clumps of cold gas across the disk or in its central regions.

\HI\ gas is detected also around the nearby galaxy IC 4200-A, resulting in a mass of $7.3\times{10^8}$ \Msun\ and a radius of $\sim$22 kpc. Although the two galaxies do not show any interaction at optical wavelength, the 21-cm picture is perhaps different. The detected gas systems are extremely close and a faint plume of \HI\ appears to be connecting IC 4200 to the companion, suggesting that the two gas disks might indeed be interacting. However, the \HI\ forming the connection fits well in the velocity structure of IC 4200, as can be seen in the right panel in Fig.\ref{pa_mos}. This suggests that the gas of the plume belongs to IC 4200 and is not being significantly disturbed by the passage of IC 4200-A, which is much less massive than IC 4200 itself. The spectacular appearance of the plume of gas in Fig.\ref{hitot} could be caused by missing gas with density lower than our detection limit in the neighbouring regions. Our data are therefore not conclusive on the actual interaction between the two galaxies.

\subsection{Radio continuum}
\label{cont_res}

\begin{figure}
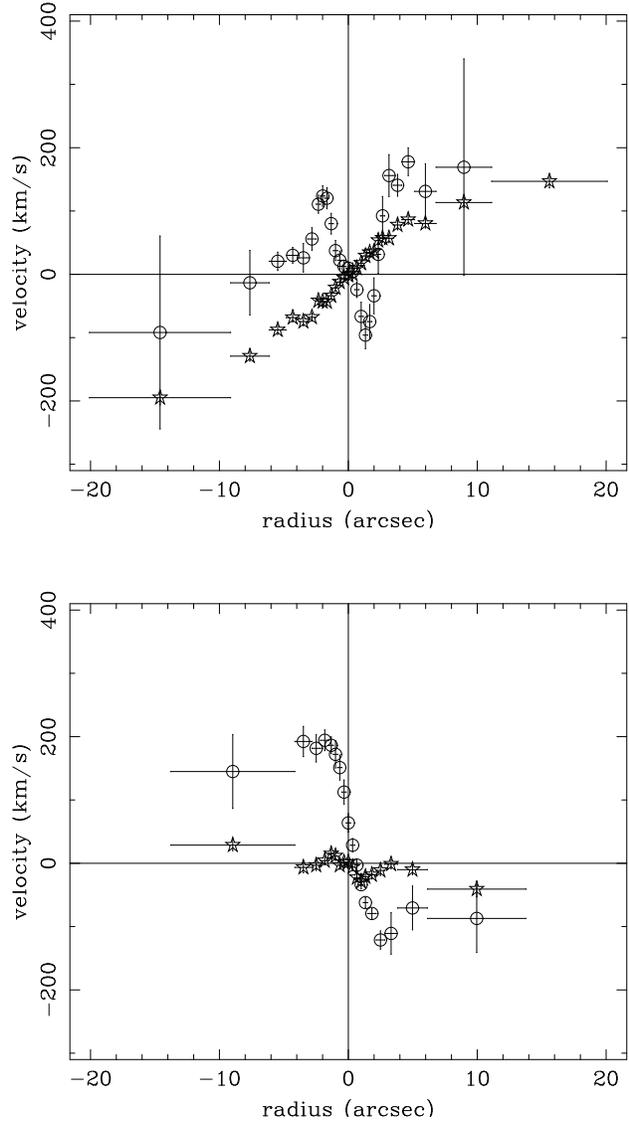

\includegraphics[width=7cm,angle=270]{kinMJnew.eps}

\vspace{0.8cm}

\includegraphics[width=7cm,angle=270]{kinminew.eps}
\caption{ Plot of the stellar and ionised-gas velocity measured from the $B_{30}$'s along the optical major (top panel) and minor (bottom panel) axis. The velocity of the ionised gas is measured from H$\alpha$ emission line.  The circles represent the ionised gas points. The horizontal bars represent the width of each bin. In order to gain S/N, the outer bins are very wide, going as far as 20 arcsec from the centre along the major axis and 14 arcsec along the minor axis. In the last bin along the major axis no reliable measurement of the gas velocity was possible. The origin of the velocity axis is the stellar systemic velocity.}
\label{kin}
\end{figure}

A unresolved source with 1.4 GHz continuum flux density of 11.5 mJy is associated to IC 4200. Previous results (strong [\OII] emission line and 60 $\mu$m detection) suggest that IC 4200 hosts some activity (star formation, nuclear activity) and therefore a non-zero 1.4 GHz flux is not surprising. To understand its origin we compare it with the emission at 60 $\mu$m (which has a flux density of 0.17 Jy). In star forming galaxies radio and FIR luminosities are tightly correlated. Using the radio-FIR relation of Yun et al.\ (2001), we find that the 1.4 GHz flux density expected from the 60 $\mu$m emission is 1.23 mJy. There is therefore a radio excess of a factor of ten, $\sim$6 times the scatter of the radio-FIR relation. This result strongly suggests that physical processes other than star formation contribute to the radio continuum (e.g. a weak AGN); 
however, it still allows some star formation to be occurring. We derive an upper limit on the star formation rate (SFR) using the FIR luminosity of IC 4200 and applying the result of Kewley et al.\ (2002). The resulting estimate is SFR=0.26 \Msun/yr.

As mentioned, IC 4200-A is also detected in the radio continuum, its flux density being 1.5 mJy. The 60 $\mu$m flux density of IC 4200-A is 0.4 Jy, which implies a radio excess of a factor of two (only 1.5 times the scatter of the relation). IC 4200-A seems to lie on the normal radio-FIR relation for star forming galaxies, with a SFR of $\sim$0.34 \Msun/yr.

\subsection{Stellar and ionised gas kinematics}

\begin{figure}
\includegraphics[width=8cm,angle=0]{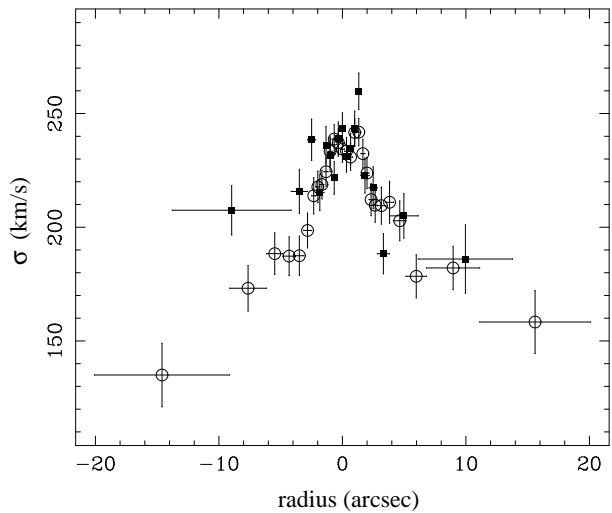}
\caption{Plot of the stellar velocity dispersion  measured from the $B_{30}$'s.   Open circles correspond to points on the major axis, filled squares correspond to the minor axis. The horizontal bars represent the width of each bin.}
\label{sigma}
\end{figure}

We sampled the kinematics of stars and ionised gas within 20 and 14 arcsec along the major and minor axis respectively \footnote{These are the $V$-band half-light radii along major and minor axis, obtained as a part of our analysis of the morphology of IC 4200 described in Sect.\ref{imag_res}.}. Fig. \ref{kin} shows stellar and gas velocities. The gas velocity is the one derived from the H$\alpha$ emission line. The obvious and main result is that stars and ionised gas have decoupled kinematics. Along the major axis the stars show $\sim$180 km/s rotation at the outer radius. In the inner 1.5-2 arcsec ($\sim$500 pc) the gas is counter-rotating compared to the stars, with the velocity steeply rising to $\sim$100 km/s. Along the minor axis, very little rotation is observed in the stellar phase, for which the velocity gradient changes sign from the inner $\sim$2 arcsec to the outer regions, and becomes again negative when looking at the outer bins $\sim$10 arcsec (2.7 kpc) from the centre. On the other hand, the gas is clearly rotating around a systemic velocity of $\sim$50 km/s and with a maximum velocity of $\sim$150 km/s.

The kinematics of the gas is certainly not consistent with coplanar circular motion, but the available data do not allow a clear interpretation. In any case, it is interesting to note that the behaviour of the ionised gas along the optical minor axis could be seen as a continuation of the velocity profile of the atomic gas in the inner region of the \HI\ disk (the left panel of Fig.\ref{pa_mos} gives a good representation of the \HI\ velocity curve along the optical minor axis). However, the difference in resolution (the kinematics of the ionised gas is sampled in a region which is roughly one tenth of the radio beam) does not allow us to further investigate this connection.

Fig.\ref{sigma} shows the stellar velocity dispersion profiles along the two optical axes. The central dispersion is of $\sim$240 km/s and decreases to 150 km/s in the outer bins along the major axis. It is interesting to compare this result with the rotation of the \HI\ disk. The latter could be modelled with a flat rotation of 320 km/s, implying a rising mass-to-light ratio and, with the assumption of spherical symmetry, an isothermal density profile ($\rho \propto r^{-2}$). For an isothermal sphere the ratio between the rotational velocity and the central velocity dispersion is $\sqrt{2}$. Interestingly, the ratio between the modelled rotational velocity and the observed line-of-sight velocity dispersion is 320/240=1.33, in reasonable agreement with the isothermal case.

\subsection{Gas ionisation}

\begin{figure}
\includegraphics[width=8cm]{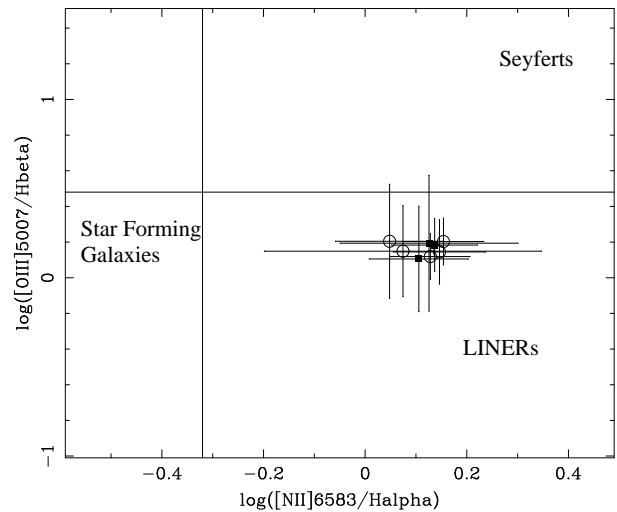}
\caption{Diagnostic diagram for the gas ionisation  derived from the $B_{80}$'s.  Open circles correspond to points on the major axis, filled squares correspond to the minor axis. Taking into account the error bars, all the points are roughly coincident; therefore individual bins were not labelled. Galaxies characterised by different ionisation mechanisms fall in different regions of the plot, as schematically indicated in this diagram (see for example Kauffmann et al.\ 2003 and references therein).}
\label{bpt}
\end{figure}

Scaling the H$\alpha$ luminosity measured along the major and minor axis, we estimate a total H$\alpha$ luminosity of 2.2$\times{10^{41}}$ ergs/s within one effective radius. To understand what causes the ionisation of the gas we use the diagnostic diagram introduced by Baldwin, Phillips \& Terlevich (1981), as shown in Fig.\ref{bpt}. Points belong to both major and minor axis and correspond to the radial bins listed in Table \ref{stellarpopresult}. The line ratios fall at all radii in the region typical of LINERs. LINERs spectra are not well understood and are commonly believed to be the product of different processes at work simultaneously: e.g., star formation, nuclear activity, shocks, old hot AGB stars.  Given our 1.4 GHz continuum result, it is not surprising to find that the ionisation of the gas cannot be explained in terms of star formation only. However, it is hard to understand which processes are going on in IC 4200. 

It is interesting that the nature of the ionisation remains the one of a LINER across the whole radial range sampled ($\sim$1.5 kpc from the centre). This suggests that diffuse shocks might be playing an important role (shocks were proposed as an explanation of LINERs spectra for example by Dopita \& Sutherland 1995). Shocks could be caused in IC 4200 by continuing gas inflow from the extended \HI\ disk or by the propagation of a jet produced by a weak AGN in the centre of the galaxy. The AGN could contribute to the observed 1.4 GHz emission, but a radio jet should produce a difference in the emission-line ratios along the two axes, which is not observed. The kinematics of the ionised gas could in principle provide a clue on whether a gas inflow or jets are present. However, although some line profiles are suspected of being composed by more than just one Gaussian, no clear evidence was observed in this sense. On the whole, we cannot reach a conclusion on the ionisation mechanism other than there is an important component not related to star formation.

\subsection{Stellar populations}
\label{spopandfrac}

\begin{figure}
\includegraphics[width=8.5cm]{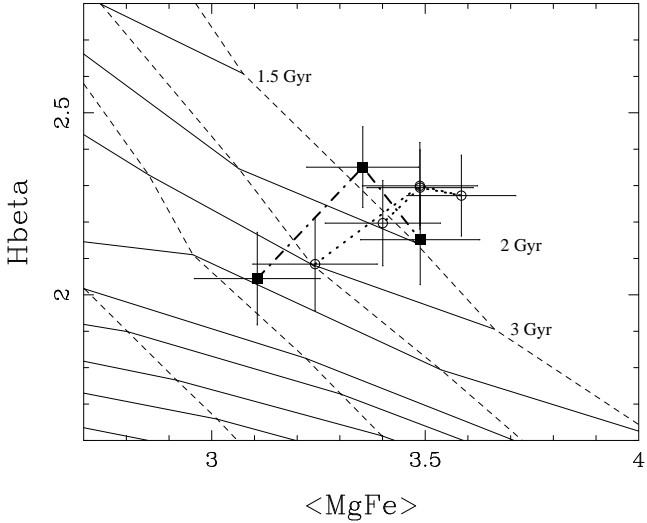}
\caption{Plot of the spectral indices  (derived from the $B_{80}$'s)  on top of the model grid of W94 (solar abundance ratios). In the plot the age runs along the $\sim$horizontal, constant-H$\beta$ solid lines, the metallicity in the perpendicular direction (dashed lines). The age of the constant-age lines is, from top to bottom, 1.5, 2, 3, 5, 8, 10, 12, 15 and 18 Gyr; metallicities [Z/H] are, --0.25, 0.0, 0.25, 0.5 ordered along increasing $<$MgFe$>$. Open circles correspond to points on the major axis and are connected to each other by a dotted line; filled squares correspond to the minor axis and are connected to each other by a dash-dotted line.}
\label{gridpoints}
\end{figure}

\begin{table}
\begin{center}
\caption{Best-fitting ages, metallicity and abundance ratios}
\label{stellarpopresult}
\begin{tabular}{c c c c c}
\hline
\hline
\noalign{\smallskip}
r (arcsec) & age (Gyr) & [Z/H] & [E/Fe] & $\chi^2$\\
\noalign{\smallskip} \hline
Major axis &              &                &                &      \\
\hline \noalign{\smallskip}
-5.5,-1.8  &  3.5$\pm$1.4 &  0.22$\pm$0.12 &  0.19$\pm$0.05 & 0.30 \\
-1.8,-0.5  &  1.6$\pm$0.3 &  0.70$\pm$0.17 &  0.28$\pm$0.05 & 1.67 \\
-0.5,+0.5  &  1.6$\pm$0.3 &  0.79$\pm$0.16 &  0.29$\pm$0.05 & 1.73 \\
+0.5,+1.8  &  1.7$\pm$0.3 &  0.69$\pm$0.16 &  0.29$\pm$0.05 & 1.93 \\
+1.8,+5.5  &  2.0$\pm$0.5 &  0.50$\pm$0.15 &  0.15$\pm$0.05 & 1.74 \\
\noalign{\smallskip} \hline
Minor axis &              &                &                &      \\
\hline \noalign{\smallskip}
-5.8,-0.8  &  5.4$\pm$2.3 &  0.05$\pm$0.12 &  0.20$\pm$0.05 & 0.04 \\
-0.8,+0.8  &  1.8$\pm$0.3 &  0.55$\pm$0.15 &  0.30$\pm$0.05 & 0.38 \\
+0.8,+5.8  &  2.2$\pm$0.6 &  0.50$\pm$0.14 &  0.24$\pm$0.05 & 0.83 \\
\noalign{\smallskip}
\hline
\end{tabular}
\end{center}
The upper part of the table corresponds to the optical major axis, the lower part to the minor axis.
\end{table}

Fig.\ref{gridpoints} shows the comparison of the measured Lick/IDS line-strength indices with Worthey (1994, hereafter W94) models on the plane [H$\beta$,$<$MgFe$>$]\footnote{$<$MgFe$>$ is defined as $<$MgFe$>$=$\sqrt{ \rm Mgb\times{<Fe>}}$, where $<$Fe$>$= =(Fe5270+Fe5335)/2}. Points correspond to the bins listed in Table \ref{stellarpopresult}. On the plane of Fig.\ref{gridpoints} age and metallicity of the models are efficiently decoupled (especially at metallicities around solar or higher), the age being well traced by H$\beta$ and the metallicity by $<$MgFe$>$. From this figure it appears that the SSP-equivalent age is $\sim$1.5 Gyr in the centre of the galaxy and increases to $\sim$3 Gyr in the outer bins.

H$\beta$, Mg$b$, Fe5270 and Fe5335 line-strength indices were used in a 4-dimensional space to more precisely derive the most likely SSP-equivalent age, metallicity and abundance ratios by comparison with W94 models extended to non-solar abundance ratios (see Trager et al.\ 2005 and references therein). The result of this analysis confirms the indications derived from Fig.\ref{gridpoints} (see Table \ref{stellarpopresult}). This implies that there is a substantial stellar component 1.5 Gyr old or younger in the centre of the galaxy.

As emphasised by Fig.\ref{gridpoints}, H$\beta$-absorption-line strength plays a crucial role when measuring the SSP-equivalent age of stellar populations. In our case, this line was studied after removing the ionised gas emission. To test the reliability of this method we compared these results with another method used by previous authors. Namely, we derived H$\beta$ absorption strength by measuring it from the original $B_{80}$'s (i.e. without subtracting the fitted ionised gas emission spectrum) and then correcting it by an additive term proportional to the equivalent width of [\OIII]5007 emission line by a factor $g$. According to Trager et al.\ (2000), $g$ varies from galaxy to galaxy within the range 0.33-1.25 and is on average 0.6. In our case the best match between the two different techniques is obtained for $g$=0.65, which fits well with previous studies.

We performed a further check on our results by studying the stellar populations with higher-order Balmer lines, namely H$\gamma$ and H$\delta$. These lines are less affected than H$\beta$ by emission contamination, and therefore provide a good test on the robustness of our emission-correction. We used the indices H$\gamma_{\rm A}$, H$\gamma_{\rm F}$, H$\delta_{\rm A}$ and H$\delta_{\rm F}$ as defined in Table \ref{ind_def}. We find that ages determined from H$\gamma_{\rm A}$ and H$\delta_{\rm A}$ (measured from the $S_{80}$'s) are in agreement with the ones determined using H$\beta$. There are systematic deviations, H$\gamma_{\rm A}$- and H$\delta_{\rm A}$-based ages being respectively lower and higher than the H$\beta$-based ones; however, these differences are comprised within the errors. On the other hand, the ages determined from H$\gamma_{\rm F}$ and H$\delta_{\rm F}$ are always higher than the ones determined from H$\beta$ by 2 Gyr or more. This discrepancy was noted already by Thomas, Maraston \& Korn (2004). In their Fig.2 one can see that ages determined from H$\gamma_{\rm A}$ are systematically lower than the ones determined from H$\gamma_{\rm F}$ and that H$\beta$-based measurements lie somewhere in between them. We finally note that we obtain the same relative ages, metallicities, and abundance ratios (for all the Balmer lines mentioned) by comparing our line strength indices to the Thomas, Maraston \& Bender (2003) models instead of the W94 ones.

We can therefore robustly conclude that the stellar populations in IC 4200 has a SSP-equivalent age of 1.5 Gyr in the centre, and that this rises to 3 Gyr in the outer region where it could be reliably determined (between $\sim r_e$/4 and $\sim r_e$/3).

\subsection{Optical Imaging}
\label{imag_res}

\begin{figure}
\includegraphics[width=8cm,angle=270]{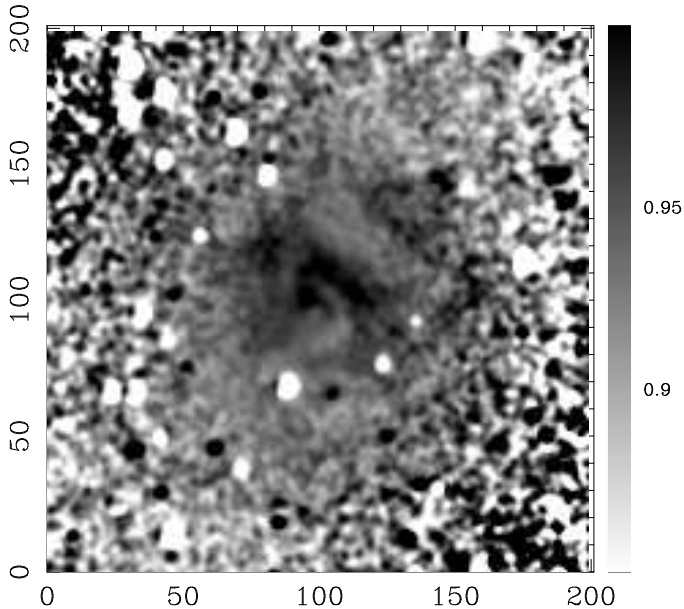}
\caption{$V-R$ colour image of the central 66$\times$66 arcsec$^2$. In the image north is up, east is left. The grey scale is in magnitudes. Axis units are in pixels, with 1 pixel=0.332 arcsec.}
\label{colimage}
\end{figure}

\begin{table}
\begin{center}
\caption{GALFIT-morphology of IC 4200 ($R$ band)}
\label{galfit}
\begin{tabular}{c c c c c c c}
\hline
\hline
\noalign{\smallskip}
           &           &   scale   & Sersic   & Axis        &       \\
Model      & Magnitude & (arcsec)  & Exponent & Ratio       & $PA$  \\
\noalign{\smallskip}
Sersic     & 11.77     & 19.2      & 4.03     & 0.696       &--23.9 \\
\noalign{\smallskip}
S+d: bulge & 12.88     & 4.6       & 1.98     & 0.746       &--22.6 \\
S+d: disk  & 12.35     & 21.5      &  1       & 0.590       &--25.5 \\
\noalign{\smallskip}
\hline
\end{tabular}
\end{center}

The $\chi^2$ of the fit is $\sim$5 for the Sersic model, $\sim$2 for the Sersic+disk model.
\end{table}

Fig.\ref{colimage} shows the $V-R$ colour image of IC 4200. This was derived after smoothing the $V$ and $R$ images with the same Gaussian filter of width $\sim$0.7 arcsec. The central region shows the presence of dust in a spiral-like pattern. This implies the existence of a small disk in the inner 5-10 arcsec. Although less clearly, the same pattern can be seen in the un-sharp-masked images in $V$ and $R$ band.

The $V$- and $R$-band morphology of IC 4200 was studied by fitting the images in the two bands with a Sersic and a Sersic+disk profile using GALFIT (Peng et al.\ 2002). The fit was performed masking out the stars as described in Sect.\ref{optobs}. In both bands the Sersic+disk profile gives a slightly better fit, but the two models are very similar within the noise of the image. Relevant fitted quantities are listed for the $R$ band in Table \ref{galfit} (the result is the same in $V$ band). The total magnitudes and colour derived from the models are in agreement with what determined manually (V=11.59, R=11.01, $V-R$=0.58 after correction for the reddening).

\begin{figure}
\includegraphics[width=9cm,angle=270]{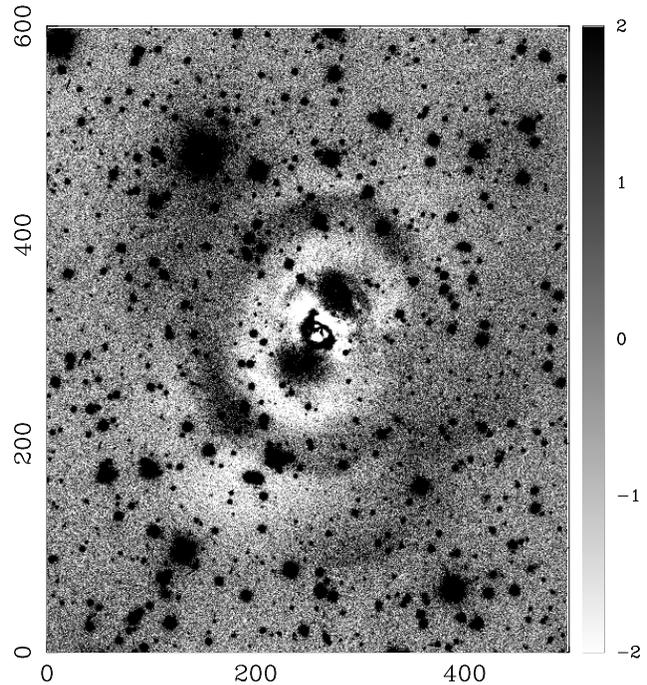}
\caption{$R$-band residual image after fitting a Sersic+disk model to IC 4200. In the image, north is up, east is left. The grey scale is in units of the rms noise. Axis units are in pixels, with 1 pixel=0.332 arcsec.}
\label{Rres}
\end{figure}

Fig.\ref{Rres} shows the $R$-band residual image obtained after subtraction of the best fitting Sersic+disk model. The general morphology of the galaxy clearly deviates from the one of a simple model. Excess light is visible $\sim$30 arcsec from the centre and is distributed in a ring-like shape. Furthermore, a faint broad shell is seen in the south part of the image. This is very likely to be the shell found by Malin \& Carter (1983). The same shell is observed in the $V$ band. The residual obtained by subtracting the best-fitting Sersic model is similar to the one of Fig.\ref{Rres}. The main difference is that the ring-like feature is much brighter and broader, especially along the optical major axis; on the other hand, the fit is slightly better in the region outside the ring.

\section{Discussion}

The results described in the last section reveal the complexity of IC 4200. The galaxy hosts a huge reservoir of \HI\ gas spread over a very extended, warped disk. This gas seems to be triggering important processes in the stellar body: optical spectroscopy suggests recent star formation episodes and reveals ubiquitous LINER-like activity, and 1.4 GHz continuum excess is observed. Furthermore, ionised gas and stars are kinematically decoupled and a disk is possibly present in the inner part of the galaxy. Finally, a stellar ring and shells are detected in the outer regions. In this section we attempt to derive a global picture of the nature and recent history of the galaxy that can explain all of the observed properties at the same time.

\subsection{Origin of the \HI\ disk: merger or IGM accretion?}
\label{discussiongas}

The \HI\ morphology and velocity structure can help placing a constraint on the age of the gas disk. The kinematics is that of a fairly regularly rotating, warped disk. This regularity suggests that the gas had time to complete at least 1-2 orbits all the way to the outer regions, where the rotational period is $\sim$1 Gyr, obtained using the model rotational velocity of 320 km/s at a radius of 60 kpc. In this time the gas can complete twice as many orbits in the inner 30 kpc, justifying the regular and symmetric appearance of the isophotes within the inner $\sim$3 beams. Therefore we conclude that the disk has been in place for at least 1-2 Gyr. No major encounters can have occurred more recently than this time.

IC 4200 contains 1.7 times more atomic gas than the Milky Way although its $B$-luminosity is only twice that of our galaxy and it belongs to an intrinsically gas-poor morphological type \footnote{The Milky Way has $L_B=2.3\times{10^{10}} L_{\odot}$ (Van den Bergh 1988).}. It is natural to ask where this exceptionally large amount of gas comes from. Three scenarios are possible: merger origin, accretion from a satellite galaxy, or accretion from the IGM.

A merger seems a plausible explanation. Barnes (2002, hereafter B02) showed that up to 60\% of the total gas of two similarly-sized merging galaxies can survive to form a very extended disk around the merger remnant. The gas disk  results from re-accretion of gas that conserved its angular momentum, forming the tidal bridges and tails observed in many merging systems, and later settled into an extended but dilute rotationally supported disk. Only a fraction of the inital gas is available for this process, while the remaining part loses its angular momentum because of shocks and gravitational torques, falling toward the centre of the remnant and there triggering star formation and probably nuclear activity.

We investigate whether it is possible to form IC 4200 within the framework of B02 simulations. The first step is to constrain the properties of the progenitor galaxies. We derive the stellar mass of IC 4200 by estimating the stellar mass-to-light ratio in $V$ band and applying the result to the observed $V$ magnitude. $M/L$=1.8 was obtained by comparison of the $V-R$ colour of IC 4200 ($V-R$=0.58) with BC03 SSP models of metallicity around solar and higher. This value implies a stellar mass of  1.1$\times$10$^{11}$\Msun, 1.9 times the stellar mass of the Milky Way \footnote{The stellar mass of the Milky Way is $6\times{10^{10}}$ \Msun\ (Bahcall \& Soneira 1980).}. Furthermore, assuming a 60\% efficency in gas disk formation, the \HI\ mass of IC 4200 (8.5$\times{10^9}$\Msun) is explained if 1.4$\times{10^{10}}$ \Msun\ of atomic gas are available during the merger -- 3 times more than the \HI\ mass of the Milky Way. In conclusion, to build the stellar and \HI\ mass of IC 4200 via an equal-mass merger, two galaxies each with 90\% of the stellar mass of the Milky Way and 1.5 times its atomic gas content are needed.

According to the simulations of B02, such systems would indeed be able to produce a very extended disk: scaling from his results, two Milky Way disk galaxies (containing $5\times{10^9}$ \Msun\ of \HI\ each in the simulations) could merge and produce a disk extended out to 30 kpc from the centre of the remnant. This result is obtained inspecting the remnant  $\sim$1 Gyr after the first encounter; however, as the time passes, more gas is re-accreted and settles into the outer parts of the disk, making it grow from the inside out and hence increasing its size. Combining this fact with the slightly higher initial gas mass required by our observations (7 rather than 5$\times{10^9}$ \Msun\ of atomic gas in each of the progenitors), it is possible to produce a disk of the mass and size of the one observed. Furthermore, in the simulations of B02 the gas of the progenitors follows the same distribution as the disk stars. In real spiral galaxies the gas reaches larger radii than stars do; including this effect would produce even larger disks around the merger remnants.

The warp of the disk can also be accommodated in the merger picture . The returning gas comes from all directions with different angular momenta and it is therefore not surprising that it does not settle into a plane. The simulations of B02 confirm that warped disks are a very common result. Our \HI\ data are therefore in a good agreement with a merger scenario for the formation of IC 4200 and would date the merger back to \emph{at least} 1-2 Gyr ago.

Are the radio data consistent with other scenarios? Accretion from another galaxy is not very likely: any accretion process capable of stripping 8.5$\times{10^9}$ \Msun\ from a passing galaxy would probably result in a major interaction. IC 4200-A itself is unlikely to be the source of all of the gas now around IC 4200; this would require that 90\% of its gas was stripped off while leaving IC 4200-A itself unperturbed. Furthermore, IC 4200-A has an \HI-mass typical of its morphological type and luminosity (derived from the Tully-Fisher relation using the observed rotation of its gas disk), so it would have been exceptionally \HI-rich if all the gas now around IC 4200 came from it.

Accretion from the IGM is another possible way of forming gas structures around galaxies. Keres et al.\ (2005; hereafter KKWD05) showed that halos can partially accrete gas from the IGM via a cold mode that does not shock heat the gas to the virial temperature of the halo but leaves it constantly below 10$^5$ K. Gas accreted via the conventional hot mode could not cool quickly enough at such low densities, and therefore could not form dilute and extended atomic gas structures like the one described in this paper. On the contrary, cold accretion would have a better chance to provide the atomic gas necessary to explain the existence of these systems.

The results of KKWD05 place important constraints on the amount of cold accretion as a function of the halo mass. We use this estimate to understand whether cold accretion can build the \HI\ disk observed around IC 4200. If the warped disk model is a fair representation of the gas structure around IC 4200, the velocity curve implies a halo mass $M_{halo}=Rv^2/G=1.7\times{10^{12}}$ \Msun\ within 60 kpc. For such a halo the fraction of gas accreted via the cold mode is quite small, going from 30\% of the accreted gas at $z$=3 to less than 5\% at $z$=0  (see Fig.6 in KKWD05). However, accretion proceeds at high rates in the young Universe, making it possible to accumulate significant masses of gas at T$<10^5$ K in an early phase. Using the results of KKWD05 for a halo of 10$^{12}$ \Msun, the average mass of gas accreted via cold mode around the central galaxy is of $1.2\times{10^{10}}$ \Msun. We obtain this result on the basis of Fig.20 in KKWD05, where the average gas accretion rate onto the central galaxy of a halo is shown for different halo masses as a function of redshift. We scale the accretion rate at each $z$ by the fraction of gas accreted via cold mode at that redshift (this fraction is obtained from Fig.6 of KKWD05). We then integrate the total mass accreted via cold mode, using the transformation $t=9.3$ Gyr/(1+z)$^{1.5}$. The result of $1.2\times{10^{10}}$ \Msun\ is close to our mass estimate for the \HI\ gas disk ($8.5\times{10^{9}}$ \Msun) \footnote{ Repeating this calculation without distinguishing between cold and hot accretion, we find that the total gaseous mass accreted is 7.5$\times{10^{10}}$ \Msun. Assuming that the hot gas is at the virial temperature of the halo ($\sim3.7\times{10^6}$ K) and that it is distributed within 60 kpc from the center, we estimate that its X-ray luminosity is $\sim5\times{10^{41}}$ ergs/s.}.

The gas accreted via cold mode has a temperature just below $10^5$ K and therefore is not in the form of \HI. However, at least a fraction of it can cool to the atomic state. Assuming a $<$10\%-in-mass enrichment from the stellar processes within the galaxy (where the metallicity is as high as 5 times solar), the accreted gas can reach $Z\sim0.1$ $Z_{\odot}$. At this metallicity and assuming a density of $\sim$10$^{-2}$ cm$^{-3}$ and no significant heating (as in IC 4200, especially at radii of tens of kpc), the cooling time is $\sim$1 Gyr. As a consequence, a significant fraction of the mass accreted via cold mode can cool to 100 K and be observed in the form of atomic gas. On the whole, considering also that KKWD05 conclusions are statistical and do not necessarily hold for any give halo, accretion from the IGM seems to be another possible way to form the observed system.

According to our calculations based on KKWD05 simulations, the largest fraction of accreted mass accumulates at higher redshifts. Assuming that accretion starts at $z$=4, 90\% of the total mass accreted via cold mode is in place at $z$=1 (this number may vary slightly depending on the environment). This implies that no substantial accretion occurred in the last $\sim$6 Gyr. If the gas around IC 4200 comes from IGM accretion, most of it has been in place for at least 6 Gyr. This is an important conclusion that has to be compared to other observational results.

\subsection{Clues from the stellar properties}

Optical observations can help to further distinguish between the merger and the IGM accretion scenarios. One result of our observations is that ionised gas and stars have different kinematics. The gas rotates in a way that may be seen as a continuation of the \HI\ velocity field, even though the spatial resolution of radio and optical data are completely different. The independent motion of gas and stars is consistent with a merger hypothesis. If hydrodynamic interactions take place gas and stars (which interact only gravitationally) can follow different behaviours and therefore end up having different kinematics. B02 simulations confirm that inner gas disks rotationally supported and decoupled from the stellar remnant can be formed in some mergers. However, the same could be said of gas accreted from the IGM, as the gas would fall in and settle in the potential independently of the underlying stellar motion.

Our results on the nature of the gas ionisation do not seem to depend on the formation scenario. LINER emission lines ratios and the 1.4 GHz radio continuum excess strongly suggest that,  perhaps along with star formation, some other activity is taking place. However, both the merger and the IGM-accretion formation scenarios could produce this situation, the relevant fact being only that the gas is present within the optical body of the galaxy, and that it might be fuelling nuclear activity and star formation, as well as producing shocks.

Two crucial results effective in understanding the formation of IC 4200 are the age of the stellar populations and the detection of shells. Star formation has occurred in the centre of the galaxy within the past 1.5 Gyr, as demonstrated by the line-strength indices presented in Sect.\ref{spopandfrac}. Assuming a two-populations model, with a young stellar population on top of a much older one, we can estimate the mass fraction of the star-formation burst that formed the young population as a function of its age.

The equivalent width EW of a given spectral feature measured within a band of width $w$ can be written as:
\begin{displaymath}
\rm EW=\it w\times \left(1-\frac{F_1+F_2}{F_{\rm c1}+F_{\rm c2}}\right),
\end{displaymath}
where $F_i$ ($i$=1,2) is the total flux coming from population $i$ and $F_{\rm c \it i}$ is the flux coming from its continuum, which is assumed to be a linear function of wavelength. Writing the flux $F_i$ of each populations in terms of its mass $M_i$ and its flux per unit mass $f_i$, and dividing and multiplying the right-hand side of the equation by the mass $M_1$ of population 1 one obtains:
\begin{displaymath}
\rm EW=\it w\times \left(1-\frac{f_1+\mu_{21}f_2}{f_{\rm c1}+\mu_{21}f_{\rm c2}}\right),
\end{displaymath}
where $\mu_{21}$ is the fraction $M_2$/$M_1$. From this expression $\mu_{21}$ can be derived as a function of the measured EW, of $w$ and of the modelled quantities $f_1$, $f_2$, EW$_1$ and EW$_2$ (EW$_1$ and EW$_2$ are the equivalent widths that would be measured from the individual spectra of population 1 and 2 respectively):
\begin{displaymath}
\mu_{21}=-\frac{f_1}{f_2}\times\frac{\rm EW_1-EW}{\rm EW_2-EW}\times\frac{w-\rm EW_2}{w-\rm EW_1}.
\end{displaymath}
From this $\mu_{2}$=$M_2$/($M_1$+$M_2$) can be easily derived.

To study the case of IC 4200 we measured the H$\beta$ line-strength index from two spectra of radial aperture equal to the $V$-band effective radius, one along the major axis and one along the minor axis (for the latter we scaled $r_e$ to the axis ratio). The index was measured in the same way as described in Sect.\ref{indices}. The average value of H$\beta$ obtained from the two axes is 2.13 \AA. The bandwidth $w$ is of 28.75 \AA\ (Table \ref{ind_def}). The other quantities that appear in the last equation were taken from the models available at Guy Worthey's web page http://astro.wsu.edu/worthey/.

We assume that population 1 and 2 are respectively 15 and 1.5 Gyr old and that both have solar metallicities. The age of the young stellar populations is the maximum allowed by the SSP-equivalent age in the centre of the galaxy (see Fig.\ref{gridpoints}). As a result we obtain $\mu_{2}$=4.1\%. Changing the assumption on the age of the old stellar populations will not change these fractions much while increasing the metallicity to [Fe/H]=0.25 brings $\mu_2$ up to 15\%. These values of $\mu_2$ are derived using 1 $r_e$ radial aperture and therefore they are fractions of the mass enclosed in one effective radius which by definition, assuming constant M/L throughout the galaxy, contains half of its stellar mass. Our result therefore implies that if the observed line strengths are caused by a burst of star formation occurred 1.5 Gyr ago, between 2.3 and 8.3$\times{10^{9}}$ \Msun\ of gas were turned into stars during this burst.

Interestingly, the mass of the star formation burst is of the order of the mass of the \HI\ disk. This is an extremely important consistency argument in favour of the merger scenario. In that picture roughly the 60\% of the gas of the progenitors had to be conserved to form the observed \HI\ disk. To have a correct balance of mass, it is necessary that the remaining 40\% is observable in IC 4200. Our stellar populations analysis is consistent with this gas mass being turned into stars during the merger. The SSP-equivalent age in the center of the galaxy provides an upper limit on the age of most massive burst of star formation induced by the merger. Such burst could occur at the time of the first encounter but also 1-2 Gyr later, during the second encounter, depending on the impact parameters (e.g. Kapferer et al.\ 2005). Therefore, the upper limit of 1.5 Gyr on the age of the burst of star formation provides an upper limit on the beginning of the merging process of $\sim$3 Gyr. Matching this indication with the lower limit on the age of the \HI\ disk (1-2 Gyr), we can conclude that if IC 4200 formed via a major merger, this must have occurred between 1 and 3 Gyr ago.

Along with the SSP-equivalent age, our analysis of line-strength indices provide estimate for the SSP-equivalent abundance ratio $\rm [E/Fe]$. We find $\rm [E/Fe]\sim$0.30 in the centre of the galaxy. Within the picture of the two-populations model, the SSP-equivalent $\rm [E/Fe]$ is determined mostly by the old and more massive population. This is a consequence of $\rm [E/Fe]$ being measured primarily on the basis of the Mgb/$\rm<Fe>$, to which cooler stars (i.e., RGB stars) contribute the most. For example, given an old population of $\rm [E/Fe]\sim$0.4 and a population 1.5 Gyr old with $\rm[E/Fe]\sim$0.1 and ten times less massive, the SSP-equivalent $\rm[E/Fe]$ is slightly above 0.3 (this result was obtained using W94 models). Therefore, our result suggests that the less massive young population lies on top of an older population formed in a short massive burst of star formation and hence characterised by high $\rm [E/Fe]$.

Can IGM accretion match the requirements imposed by stellar populations results? As mentioned, IGM accretion implies that the disk formed more than 6 Gyr ago. The accreted atomic gas might have fuelled star formation at that time, but it is not clear why such a massive burst of residual star formation (that would have consumed roughly half of the mass accreted as cold gas) should have occurred so much later, in the last 1.5 Gyr. IGM accretion is therefore hard to reconcile with the stellar population properties of IC 4200.

Some of the morphological features (the inner disk and the ring-like feature) can be explained by both merger and IGM accretion. However, the detection of shells in the images of the galaxy points toward a merger origin. It is believed that shells can form as a result of accretion of smaller satellites by bigger galaxies (e.g.\ Quinn 1984) or by equal mass mergers (Hernquist \& Spergel 1992). On the contrary, IGM accretion alone cannot explain shell formation.

In conclusion, the only formation process that can by itself explain all the observed properties is a major merger. According to the observational evidence, a merger of two galaxies each with 90\% of the Milky Way stellar mass and containing 1.5 times its gas mass occurred between 1 and 3 Gyr ago. During the merger roughly half of the atomic gas of the progenitors retained its angular momentum and was later re-accreted, settling into a 90-deg warped disk extended out to 60 kpc from the center of the galaxy. The loss of angular momentum drove the remaining gas toward the centre of the newly formed galaxy, where it triggered star formation.  This resulted also in the formation of a decoupled central component visible as a spiral-like dust pattern. The dynamically violent process produced a merger remnant with morphological disturbances that are still visible in the outer regions.  Finally, ionised gas is present in the stellar body and, as a result of the merger, is kinematically decoupled from the stars.  It is also fuelling some activity, possibly a central AGN or shocks, and maybe star formation.

We note that the non-detection of IC 4200 in the X-ray supports our conclusion on the origin of the galaxy. As suggested by Sansom et al. (2000), it probably takes several gigayears for hot gas halos to build up after a merger or a strong interaction. They reached this conclusion on the basis of the anti-correlation between X-ray excess and morphological fine structure. Given its $L_B$ (and following the mean relation between $L_B$ and $L_X$ of Sansom et al.\ 2000), IC 4200 should have been detected by X-ray surveys (e.g., $ROSAT$). The same is true if we use our estimate of $L_X$ based on the accretion of hot gas from the IGM (Sec. \ref{discussiongas}). The fact that IC 4200 has not been detected in the X-ray is therefore consistent with a major merger origin of the galaxy.

\subsection{The fate of IC 4200}

A final interesting question is what the future evolution of IC 4200 will be. Given the amount of \HI\ gas and its configuration, it is very unlikely that it will ever become a more typical gas-poor member of its morphological type. The gas seems to be in a quite regular and stable configuration, having been in place already more than 1 Gyr. If, on the other hand, the gas density increases just enough to start star formation, it is possible that IC 4200 will grow a stellar disk, becoming an early-type spiral. In this case it would be an exceptionally bulge-dominated galaxy, with M$_{bulge}$/M$_{disk}\sim$15 or more and a total stellar mass of $\sim$1.2$\times$10$^{11}$ \Msun. The mass-to-light ratio of the galaxy is currently $\sim$30 in the $V$ band and out to 60 kpc and would maintain roughly the same value in case of the formation of a stellar disk.

\section{Conclusions}

\begin{table}
\begin{center}
\caption{Final Result}
\label{conclusion}
\begin{tabular}{c|c c c}
\hline
\hline
\noalign{\smallskip}
& & Satellite & IGM \\
& Merger & Accretion & Accretion \\
\noalign{\smallskip}
\hline
\noalign{\smallskip}
\HI\ & $\surd$ & $\times$ & $\surd$ \\
stellar populations & $\surd$ & $\times$ & $\times$ \\
shells & $\surd$ & $\surd$ & $\times$ \\
\noalign{\smallskip}
\hline
\end{tabular}
\end{center}
\end{table}

We have presented the result of radio (21-cm-line emission and 1.4 GHz continuum) and optical (long-slit spectroscopy and imaging) observations of the early-type galaxy IC 4200. A total of $8.54\times{10^9}$ \Msun\ of \HI\ is detected around this object and  $7.3\times{10^8}$ \Msun\ around its companion galaxy IC 4200-A. The gas around IC 4200 can be modelled as a warped disk and its structure suggests that it has been in place for at least 1-2 Gyr. The stellar populations of IC 4200 have a 1.5 Gyr SSP-equivalent age. Ionised gas is present across the whole stellar body and is kinematically decoupled from the stars. This gas shows a spectrum characteristic of LINERs across the inner $r_e$/3. The activity is confirmed by a 1.4 GHz continuum excess compared to the 60 $\mu$m emission. An optical shell have been detected around the galaxy next to a ring-like feature of radius 30 arcsec.

Table \ref{conclusion} summarises the conclusions we derive from these observational results. The merger scenario is the only one that can explain at once all our results, while IGM or satellite accretion fail to do so. From \HI\ data we conclude that IC 4200 formation event must have occurred more than 1-2 Gyr ago. From the SSP-equivalent age of the stellar populations we derive that the formation of the galaxy has occurred via a merger less than $\sim$3 Gyr ago (taking into account a delayed burst of star formation). Matching these two results we can claim that IC 4200 formed via a merger between 1 and 3 Gyr ago. Within this interpretation of the data, two Milky-Way-like progenitors are required in order to produced the observed stellar and \HI-gas masses.

Table \ref{conclusion} stresses the power of our approach in determining the past history of galaxies. Only the combination of 21-cm data, optical imaging and stellar populations analysis allowed us to obtain a robust result. Following papers will present a similar study of a larger number of galaxies of our sample.

\begin{acknowledgements}

The authors wish to thank the anonymous referee for the constructive comments. We would like to thank Daniel Kelson for his help during the reduction of the long-slit spectroscopy data. We would also like to thank Guy Worthey for providing the material necessary for the calibration of the line-strength indices onto the Lick/IDS system. Finally, we thank Jacqueline van Gorkom and Renzo Sancisi for stimulating discussions on the origin of gas in early-type galaxies. This research has made use of HyperLeda and of the NASA/IPAC Extragalactic Database (NED) which is operated by the Jet Propulsion Laboratory, California Institute of Technology, under contract with the National Aeronautics and Space Administration. This work is based on observation with the Australia Telescope Compact Array, which is operated by the CSIRO Australia Telescope National Facility, and on observations collected at the European Southern Observatory, La Silla, Chile. 

\end{acknowledgements}

\appendix
\section{Calibration of the line-strength indices to the Lick/IDS system}
\label{Lick/IDScal}

\begin{table}
\begin{center}
\caption{MILES to Lick/IDS transformation}
\label{M2L}
\begin{tabular}{c c c c}
\hline
\hline
\noalign{\smallskip}
Index & Slope ($m$) & Intercept ($q$) & RMS \\
\noalign{\smallskip} \hline \noalign{\smallskip}
H$\beta$    & 0.97 &    0.12 & 0.27 \\
Mg$b$       & 0.88 &    0.31 & 0.41 \\
Fe5270      & 0.91 &    0.16 & 0.32 \\
Fe5335      & 0.94 &    0.23 & 0.34 \\
H$\gamma_{\rm A}$ & 0.95 &  --0.30 & 0.80 \\
H$\delta_{\rm A}$ & 0.94 &  --0.17 & 0.85 \\
H$\gamma_{\rm F}$ & 0.96 &  --0.15 & 0.46 \\
H$\delta_{\rm F}$ & 0.95 &    0.02 & 0.49 \\
\noalign{\smallskip}
\hline
\end{tabular}
\end{center}
The calibrated index is $i_{\rm Lick/IDS}=m\times{i_{\rm MILES}}+q$; the RMS is calculated on the basis of $\sim$240 stellar spectra.
\end{table}

\begin{table}
\begin{center}
\caption{INDO-US to Lick/IDS transformation}
\label{I2L}
\begin{tabular}{c c c c}
\hline
\hline
\noalign{\smallskip}
Index & Slope ($m$) & Intercept ($q$) & RMS \\
\noalign{\smallskip} \hline \noalign{\smallskip}
H$\beta$    & 1.01 &    0.13 & 0.25 \\
Mg$b$       & 0.89 &    0.32 & 0.35 \\
Fe5270      & 0.97 &    0.12 & 0.29 \\
Fe5335      & 1.01 &    0.08 & 0.32 \\
H$\gamma_{\rm A}$ & 1.01 &  --0.15 & 0.71 \\
H$\delta_{\rm A}$ & 0.98 &  --0.02 & 0.83 \\
H$\gamma_{\rm F}$ & 1.01 &  --0.10 & 0.43 \\
H$\delta_{\rm F}$ & 0.98 &    0.14 & 0.51 \\
\noalign{\smallskip}
\hline
\end{tabular}
\end{center}
The calibrated index is $i_{\rm Lick/IDS}=m\times{i_{\rm INDO-US}}+q$; the RMS is calculated on the basis of $\sim$150 stellar spectra.
\end{table}

Our analysis of the stellar populations of IC 4200 relies on the comparison of its spectral indices with the ones predicted by the stellar populations models of W94. These models are built to match age and metallicity of the stars included in the Lick/IDS library (Worthey et al.\ 1994), which have been observed in well defined instrumental conditions and are not flux-calibrated. Therefore, any comparison of the spectral indices of IC 4200 to these models makes sense only if the indices themselves are onto the Lick/IDS system. A practical way to satisfy this requirement is to observe with the same instrumental setup used for IC 4200 a sample of stars that already have indices on the Lick/IDS system. We can then measure the spectral indices of these stars from our spectra and calculate the transformation necessary to make them match their Lick/IDS-reference values. The same transformation will then be applied to the indices measured from the spectrum of IC 4200. Once this is done, the spectral indices of IC 4200 can be compared to W94 models.

\begin{figure}
\includegraphics[width=8.75cm]{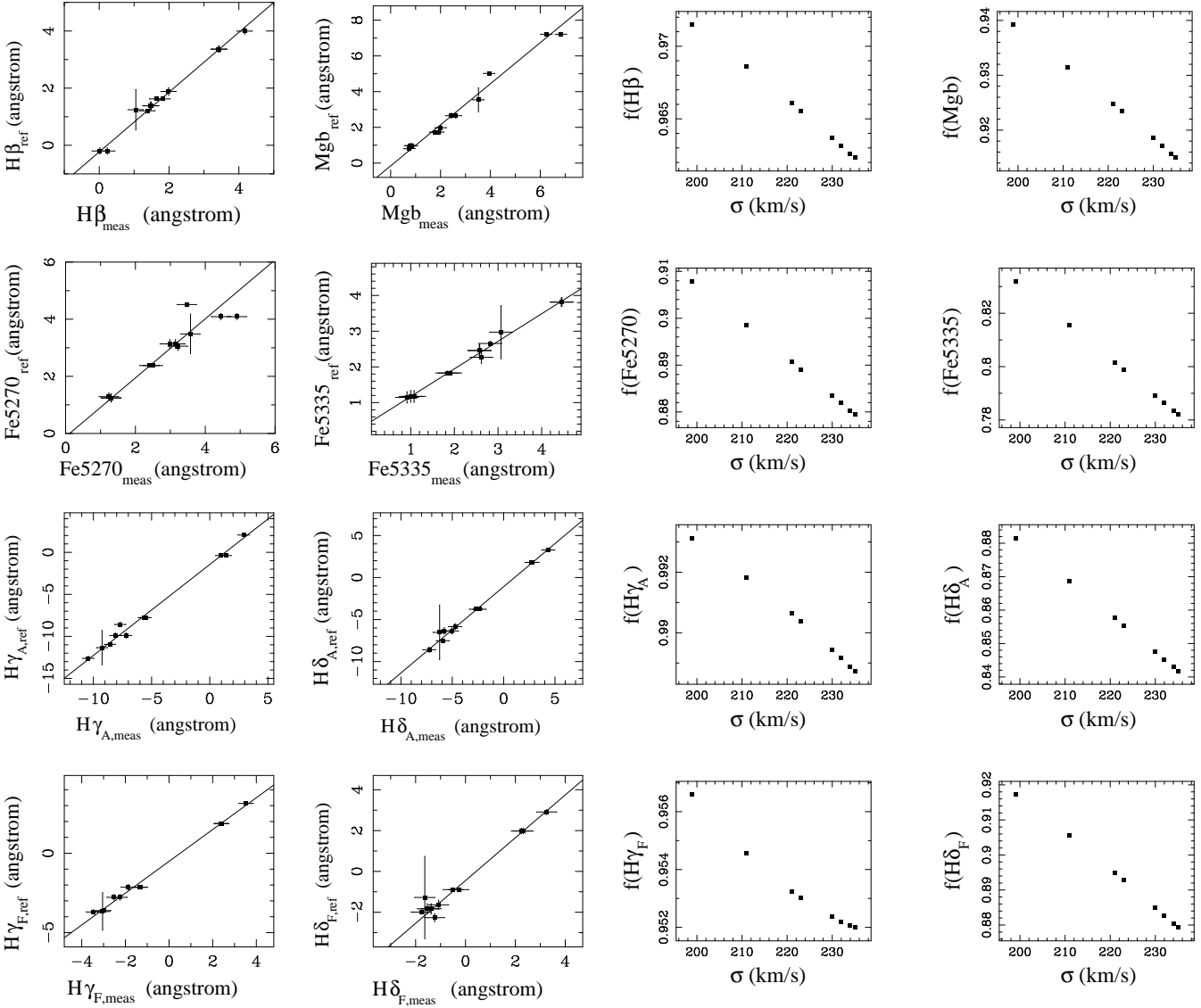}
\caption{Calibration to the Lick/IDS system. For each index the points in the plane ($i_{\rm ref},i_{\rm meas}$) are showed together with the best fitting straight line. Each point represents one of our Lick/IDS stars. The fitting was performed weighting points according to their errors in both coordinates.}
\label{lickcalib}
\end{figure}

The stars observed in order to perform the calibration are listed in Table \ref{LICKlist}. Guy Worthey kindly provided us the reference values of the spectral indices for these stars. However as indicated in Table \ref{LICKlist}, only two of them belong to the Lick/IDS library and therefore have indices already on the Lick/IDS system. Indices measured from spectra in either the MILES (S\'{a}nchez-Bl\'{a}zquez et al.\ 2006, in preparation) or the INDO-US library (Valdes et al.\ 2004) are not calibrated and had to be brought to the Lick/IDS system. We did this by means of a set of transformations also kindly provided by Guy Worthey. For each index, a linear transformation brings from the ``MILES value'' to the Lick/IDS system and another one brings from the ``INDO-US value'' to the Lick/IDS system. The coefficient of these transformations for the indices used in this paper can be found in Table \ref{M2L} and Table \ref{I2L}.

\begin{figure}
\includegraphics[width=8.75cm]{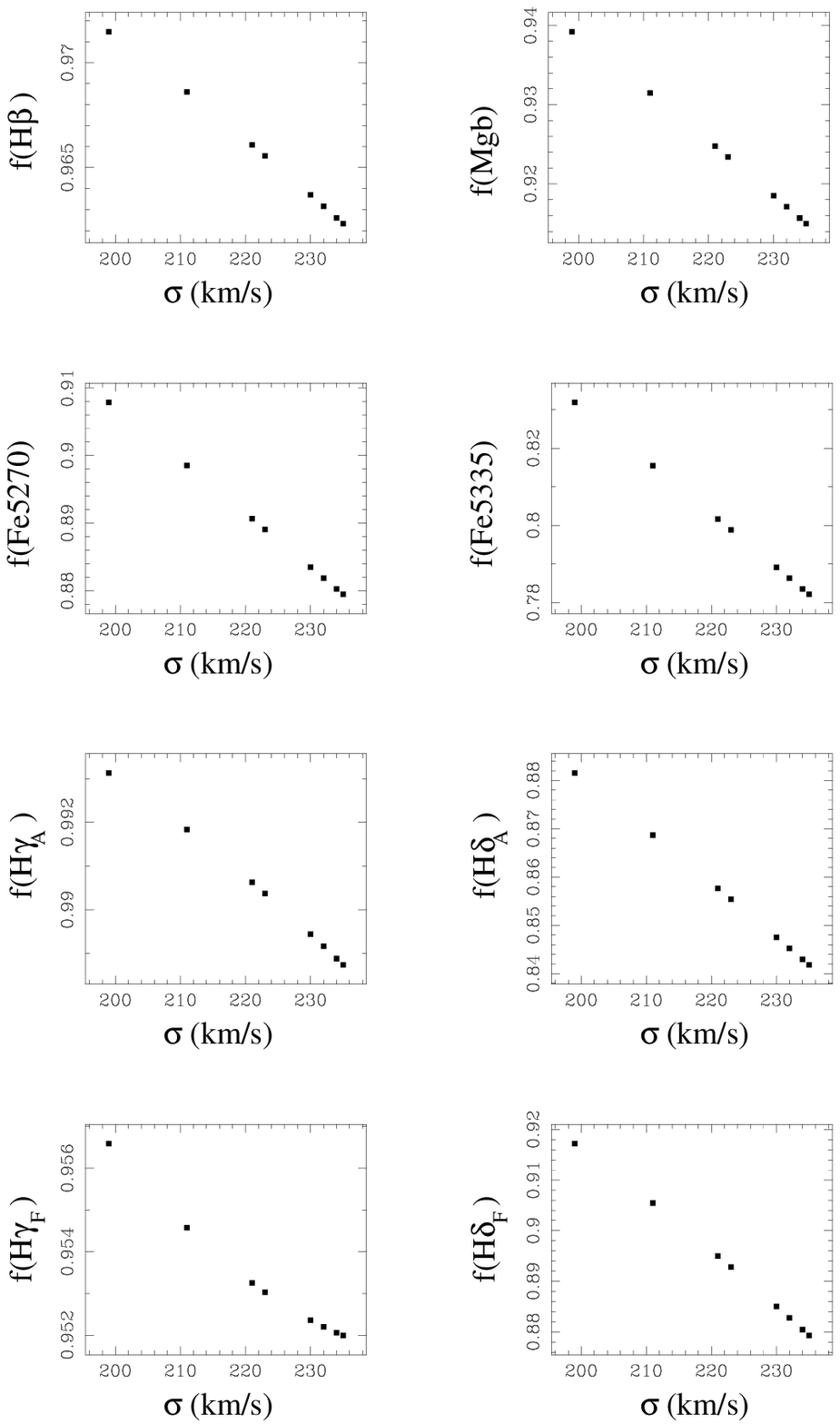}
\caption{Correction of the line-strength indices for the effect of the stellar velocity dispersion. The ratio $f$ between the measured and the corrected value is plotted versus the velocity dispersion. Each point correspond to one of the S$_{80}$'s used to study the stellar populations in IC 4200.}
\label{fplots}
\end{figure}

Once all the stars have Lick/IDS-reference indices values, the calibration of the generic index $i$ was done by:

\begin{itemize}
\item broadening our stellar spectra to match the Lick/IDS resolution
\item measuring $i$ for each of our Lick/IDS stars obtaining the values $i_{j,\rm meas}$, where $j$=$1,...,N_{\rm stars}$
\item comparing $i_{j,\rm meas}$ to the respective reference values $i_{j,\rm ref}$; by fitting a straight line to the N$_{\rm stars}$ points in the plane ($i_{\rm ref},i_{\rm meas}$) we could obtain for each index of interest the linear relation between the measured and the reference value.
\end{itemize}

Fig.\ref{lickcalib} shows the planes ($i_{\rm ref},i_{\rm meas}$) with the data points and the linear fit. The coefficient of the linear relation $i_{\rm meas}=m\cdot i_{\rm ref}+q$ are given in Table \ref{Lick_cal}.

\section{Velocity dispersion corrections of the line-strength indices}
\label{vdispcorr}

Fig.\ref{fplots} shows the factor $f$ used to correct line-strength indices for the stellar velocity dispersion as a function of the velocity dispersion for the indices used in this paper. $f$ is defined so that the corrected line-strength index is $i_{0,k}=i_k/f$ where $i_k$ is the measured value. For more details see Sect.\ref{indices}.

\section{Tables}
\label{eltables}

Tables C.1 and C.2 are available electronically at the CDS. Table C.1 contains informations on the stellar and ionised-gas kinematics along the $B_{30}$'s. Column 1 specifies the axis (either major or minor axis), Columns 2 and 3 give the lower and upper radial coordinate of the bins, Columns 4 and 5 list the stellar line-of-sight velocity and its error, Columns 5 and 6 give the stellar line-of-sight velocity dispersion ($\sigma$) and its error, Columns 7 and 8 provide the central wavelength of H$\alpha$ emission line and its error, and finally Columns 9 and 10 give the FWHM of H$\alpha$ emission line and its error.

Table C.2 lists the value of all the Lick/IDS indices fully corrected and calibrated. In this table Column 1 specifies the axis (either major or minor axis), Columns 2 and 3 give the lower and upper radial coordinate of the bins from which the indices were measured, and the following 50 columns give the value of the 25 Lick/IDS indices, each followed by its error.

\end{document}